\documentclass[aps, notitlepage, superscriptaddress, pre, reprint, longbibliography]{revtex4-1}

\usepackage{graphicx, hyperref, natbib, titlesec, paralist, amsmath, amssymb, color,array, caption, subcaption, multirow}
\def\*#1{\mathbf{#1}}

\makeatletter
\renewcommand*{\@fnsymbol}[1]{\ensuremath{\ifcase#1\or \dagger \or * \or \ddagger\or
   \mathsection\or \mathparagraph\or \|\or **\or \dagger\dagger
   \or \ddagger\ddagger \else\@ctrerr\fi}}
\makeatother

\begin{document}

\title{Obstructed swelling and fracture of hydrogels}

\author{Abigail Plummer}
\thanks{A.P. and C.A. contributed equally to this work.}
\affiliation{Princeton Center for Complex Materials, Princeton University, Princeton, NJ 08540.}

\author{Caroline Adkins}
\thanks{A.P. and C.A. contributed equally to this work.}
\affiliation{Department of Civil and Environmental Engineering, Princeton University, Princeton, NJ 08544.}

\author{Jean-Fran\c{c}ois Louf}
\affiliation{Department of Chemical and Biological Engineering, Princeton University, Princeton, NJ 08544.}
\affiliation{Department of Chemical Engineering,  Auburn University, Auburn, AL 36849.}

\author{Andrej Ko\v{s}mrlj}
\thanks{To whom correspondence may be addressed. \\Email: \tt{andrej$@$princeton.edu} or \tt{ssdatta$@$princeton.edu}.}
\affiliation{Department of Mechanical and Aerospace Engineering, Princeton University, Princeton, NJ 08544.}
\affiliation{Princeton Materials Institute, Princeton University, Princeton, NJ 08544.}

\author{Sujit S. Datta}
\thanks{To whom correspondence may be addressed. \\Email: \tt{andrej$@$princeton.edu} or \tt{ssdatta$@$princeton.edu}.}
\affiliation{Department of Chemical and Biological Engineering, Princeton University, Princeton, NJ 08544.}

\date{\today}

\begin{abstract}
Obstructions influence the growth and expansion of bodies in a wide range of settings---but isolating and understanding their impact can be difficult in complex environments. Here, we study obstructed growth/expansion in a model system accessible to experiments, simulations, and theory: hydrogels swelling around fixed cylindrical obstacles with varying geometries. When the obstacles are large and widely-spaced, hydrogels swell around them and remain intact. In contrast, our experiments reveal that when the obstacles are narrow and closely-spaced, hydrogels fracture as they swell. We use finite element simulations to map the magnitude and spatial distribution of stresses that build up during swelling at equilibrium in a 2D model, providing a route toward predicting when this phenomenon of self-fracturing is likely to arise. Applying lessons from indentation theory, poroelasticity, and nonlinear continuum mechanics, we also develop a theoretical framework for understanding how the maximum principal tensile and compressive stresses that develop during swelling are controlled by obstacle geometry and material parameters. These results thus help to shed light on the mechanical principles underlying growth/expansion in environments with obstructions.
\\ 


\end{abstract}

\maketitle
\section{Introduction}
Many growth and expansion processes are sculpted through confinement by rigid obstructions. Familiar examples include muffins rising into their characteristic shape during baking \cite{ding2019thermomechanical}, trees growing around boulders \cite{taylor2021mechanism}, and even cities expanding around inhospitable geographic features \cite{borsdorf2020urban}. Obstructed growth and expansion also play pivotal roles---both harmful and beneficial---in many practical applications. For example, excessive tissue growth around metal mesh tubes inserted into blood vessels is a common, but life-threatening, complication of stenting \cite{byrne2015stent, kuhl2007computational, cheng2021finite}; conversely, the expansion of spray foam into cracks and in between walls underlies the thermal insulation of many energy-efficient buildings \cite{cabeza2010experimental}. More broadly, obstructed growth and expansion critically influence the emergence of form and function across diverse non-living and living systems, ranging from hydrogels added to soil for water retention to  biofilms and biological tissues in complex environments~\cite{louf_under_2021, beebe2000functional, alben2022packing, chu2018self, streichan2014spatial, bengough1997biophysical, alric2022macromolecular}. Therefore, we ask: are there general principles that dictate how obstructions influence growth and expansion? And if so, how do we discover them?

\begin{figure*}[htp]
    \includegraphics[width=\linewidth]{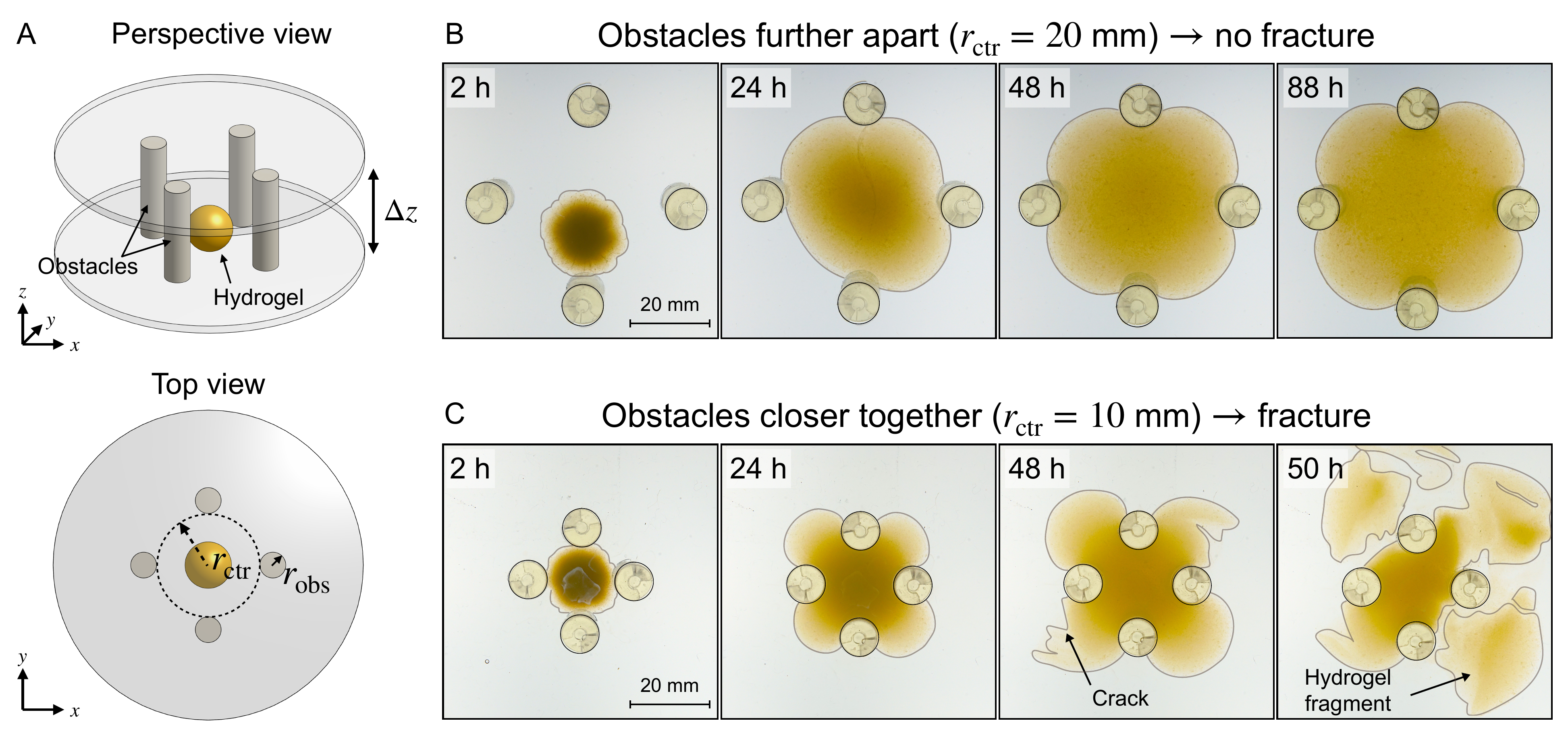}
    \caption{\textbf{Hydrogels swelling around obstacles remain intact at equilibrium when obstacles are further apart, but fracture when obstacles are closer together.} (A) Schematic of the experimental platform, showing a hydrogel (yellow) surrounded by cylindrical obstacles with radius $r_{\rm{obs}}$ separated according to $r_{\rm{ctr}}$ and confined vertically by parallel plates separated by $\Delta z$. (B-C) Top view images taken over the course of swelling, with $r_{\rm{obs}}=5~\rm{mm}$. The approximate borders of the obstacles and hydrogels are outlined for clarity.  (B)~When the obstacles are further apart ($r_{\rm{ctr}}=20~\rm{mm}$), a hydrogel reaches a four-lobed clover-like shape at equilibrium. (C)~When the obstacles are closer together ($r_{\rm{ctr}}=10~\rm{mm}$), cracks appear at the surface of the hydrogel as it swells, driving the repeated production of discrete fragments. 
}
    \label{fig:schematic}
\end{figure*}

When the growth/expansion of a body is resisted by surrounding obstructions, large and spatially-nonuniform
stresses can develop, influencing subsequent growth/expansion in turn~\cite{ambrosi2019growth, goriely}. Being able to understand the distribution and magnitude of these stresses is thus a necessary step in the development of widely-applicable, predictive models of growth and expansion. However, model systems in which the coupling between growth and stress can be systematically studied in structured environments are scarce. Here, we address this issue using studies of spherical hydrogel beads swelling in 3D-printed obstacle arrays with defined geometries. Hydrogels are cross-linked networks of hydrophilic polymers that can absorb large amounts of water and swell while still retaining integrity. As a result, they are extensively used in biomedical, environmental, and manufacturing
applications, and have well-characterized and highly-tunable properties such as their degree of swelling and elasticity \cite{zhang2017advances, oyen2014mechanical, deligkaris2010hydrogel, guo2020hydrogels}. Indeed, the comprehensive theoretical literature on hydrogel swelling makes computations of shape and internal stresses accessible for a variety of boundary conditions. While much work has focused on the case of a hydrogel swelling while adhered to another material~\cite{tanaka1987mechanical, hong2008theory, hong2009inhomogeneous,kang2010variational,  marcombe2010theory, amar2010swelling,dervaux2012mechanical, kim2010dynamic, cho2019crack, dortdivanlioglu2021swelling}, non-adhered swelling around obstructions has received more limited attention~\cite{hui2006contact, hong2008theory, hu2010using, bouklasfem, levin2023swelling, he2012modeling}, often in the context of indentation testing. 

Our study extends this body of work. First, we use experiments to directly visualize how hydrogel swelling is altered by obstacles of systematically-varying geometries. When obstacles are positioned further apart, the hydrogel swells through the spaces between them and maintains its integrity. By contrast, when the obstacles are closer together, we observe a dramatic phenomenon: the hydrogel fractures, repeatedly tearing itself apart as it swells! We then use theory and simulations to rationalize these observations, and importantly, quantify and understand the distribution of stresses. Taken together, our work provides a prototypical example of obstructed growth/expansion and uncovers complex swelling behaviors and mechanical instabilities that can result during this process---highlighting the rich physics waiting to be explored in this area of soft mechanics.

\section{Experiments}
Our experimental platform is schematized in Fig. \ref{fig:schematic}A and detailed in \textit{Materials and Methods}. To define the obstacles, we 3D-print four rigid cylindrical columns of radius $r_{\rm{obs}}$ to be placed an equal distance $r_{\rm{ctr}}$ from a central point; in the experiments, we vary $r_{\rm{obs}}$ and $r_{\rm{ctr}}$ between $2.5-15$~mm and $7.5-25$~mm, respectively. The cylinders are securely attached to horizontal, parallel laser-engraved acrylic plates spaced vertically by a fixed amount, $\Delta z=40$~mm (see SI Sec. B for a discussion on the selection of this value). Importantly, these plates are transparent, permitting direct visualization of a hydrogel as it swells between the cylindrical obstacles and parallel plates. Hence, at the beginning of each experiment, we place a spherical polyacrylamide hydrogel bead of initial radius $\sim 6$~mm (initial state characterized in SI Sec. A) in the center and submerge the entire apparatus in a bath of ultrapure milli-Q water---thereby initiating swelling, which we track using a camera focused on the top plate. 

We first examine the case of obstacles that are spaced further apart. As the hydrogel swells, it contacts the top and bottom plates, as well as the cylinders, and continues to swell through the space between them (Fig.~\ref{fig:schematic}B, Movie S1). It eventually reaches an unchanging four-lobed equilibrium shape. As the hydrogel swells, the yellow dye fixed in its polymer network becomes more dilute; thus, the color intensity serves as a proxy for the local polymer concentration. However, the deformed 3D shape of the hydrogel makes it challenging to extract quantitative information about relative expansion from a purely top-down view. We leave this effort for future work.

The case of closer-spaced obstacles of the same size is dramatically different. We observe similar behavior to the previous case of less confinement at early times: the hydrogel contacts the surrounding surfaces and swells through the space between them. As these lobes continue to swell, however, cracks abruptly form at the hydrogel surface (48~h in Fig.~\ref{fig:schematic}C), reflecting the development of large stresses during obstructed swelling. Remarkably, this sequence then continues, resulting in elaborate, multi-step fracturing of the hydrogel as it swells, repeatedly ejecting fragments of the hydrogel outward (Fig.~\ref{fig:schematic}C, Movie S2).  The fracturing process is dynamic: As cracks propagate through compressed areas, releasing stresses, those regions are then able to increase their solvent content and swell. The process eventually stops as the central hydrogel body reaches its final equilibrium degree of swelling. Another representative example of a fracturing hydrogel with a different obstacle geometry is shown in Movie S3. The fracturing process varies significantly between samples, reflecting the acute sensitivity of crack formation and propagation on the random imperfections in the hydrogel and the complex topography arising from previous fracturing~\cite{gdoutos2020fracture,wang2022hidden, li2023crack}.

\begin{figure}
    \includegraphics[width=\columnwidth]{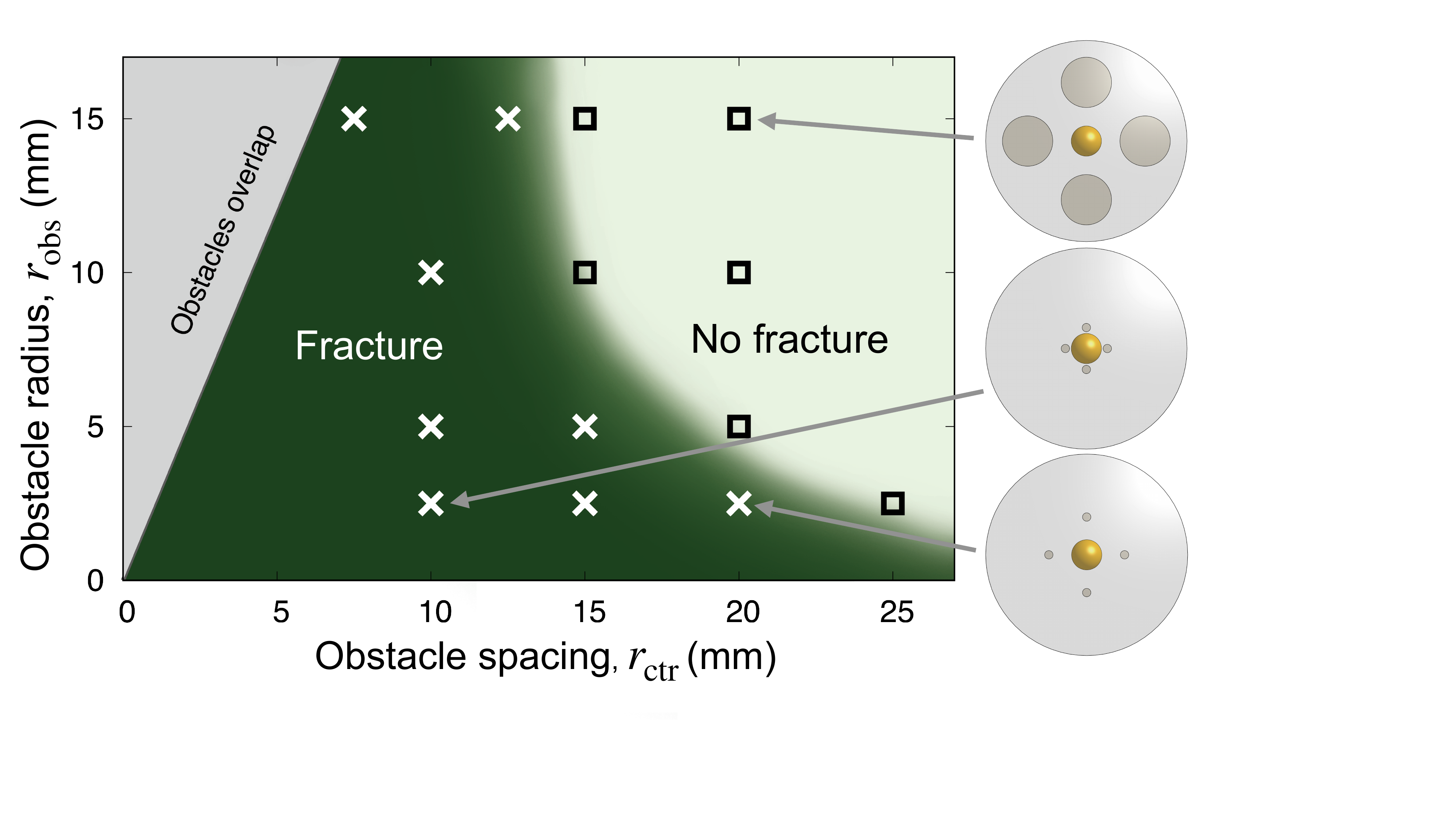}
    \caption{\textbf{Experiments reveal a hydrogel fracture threshold that depends on obstacle radius and spacing.} Each $\square$ indicates an experiment in which the hydrogel reached an intact equilibrium shape as in Fig.~\ref{fig:schematic}B, while each $\times$ indicates an experiment that resulted in fracture as in Fig.~\ref{fig:schematic}C. Schematics show a top view of the obstacle geometries for the indicated points. The grey excluded region in the top left shows parameters for which the obstacles would overlap. The green background shading provides a guide to the eye.}
    \label{fig:threshold}
\end{figure}

These observations suggest that, when appropriately obstructed, a growing/expanding body can tear itself apart. To characterize the dependence of this phenomenon on confinement, we repeat these experiments with obstacles of varying sizes $r_{\rm{obs}}$ and spacings $r_{\rm{ctr}}$. We observe a similar phenomenon in all cases, as summarized in the state diagram in Fig. \ref{fig:threshold}. When the obstacles are far apart, the hydrogel swells and retains its integrity ($\square$), while when the obstacles are closer i.e., $r_{\rm{ctr}}$ is below a threshold value, the hydrogel repeatedly self-fractures as it swells ($\times$). The fracturing threshold also depends on the size of the obstacles: For a given $r_{\rm{ctr}}$, fracturing occurs \textit{below} a threshold value of $r_{\rm{obs}}$. The finding that smaller obstacles promote fracture is reminiscent of wire cutting tests in fracture mechanics, in which thin wires pushed into soft materials induce cutting~\cite{baldi2012measurement, kamyab1998cutting}. The two geometric parameters $r_{\rm{ctr}}$ and $r_{\rm{obs}}$ thus delineate a boundary between swelling without fracture and obstructed swelling causing fracture, as shown by the light and dark green regions, respectively, in Fig. \ref{fig:threshold}. Intuitively, this non-trivial dependence of the onset of fracturing on both obstacle spacing and size reflects the balance between the osmotic pressure driving swelling---which is set by the difference between the solvent chemical potential in the unswollen hydrogel compared to at its boundary---and the tensile stresses that develop in the hydrogel as it swells against narrow obstacles~\cite{louf_poroelastic_2021, louf_under_2021}. We theoretically model and computationally quantify this balance in the following sections.

To demonstrate the generality of this phenomenon, we repeat our experiments in fully 3D granular packings, which more closely mimic the obstructions experienced by hydrogels in many applications \cite{misiewicz2022characteristics, krafcik2017improved, kapur1996hydrodynamic, dolega2017cell, lee2019dispersible, woodhouse1991effect, demitri2013potential, guilherme2015superabsorbent, lejcus2018swelling, wei2013effect}. Our previous experiments \cite{louf_under_2021} using this platform used a solvent that promoted only slight hydrogel swelling, and therefore accessed the case of weak confinement; however, repeating these experiments with a solvent that promotes more hydrogel swelling indeed gives rise to self-fracturing, as shown in Fig. S4 (SI Sec. C). Thus, this phenomenon of obstructed growth/expansion causing fracture arises not just in idealized geometries, but also in more realistic complex spaces. 

\section{Theory and simulations}
What are the essential physical principles that govern this phenomenon? To address this question, we develop finite element simulations that incorporate the energetic penalty of contacting obstructions into a model of hydrogel swelling based on classic Flory-Rehner theory. Our goal is to better understand the complex distribution of stresses that arises during obstructed swelling, not to quantitatively capture all the features of the experiments; as such, we use a simplified two-dimensional (2D) model that permits straightforward visualization of stresses and thereby enables us to develop an intuitive and analytic description of the underlying physics. The dimensional reduction is discussed at length in SI Sec. D.

We describe stretching of the hydrogel polymer network and solvent-polymer interactions with the commonly-used free energy density \cite{floryrehner, flory, hong2008theory, hong2009inhomogeneous, kang2010variational, bouklasfem}:
\begin{align}
    \frac{F_{\rm{en}}}{A_0 k_B T}&=\frac{n_p}{2}  \left( F_{iK} F_{iK} - 2 - 2 \ln(\det(\*F))\right) \nonumber \\&+\frac{1}{\Omega} \left(\Omega C \ln \left( \frac{\Omega C}{1+\Omega C} \right) + \chi \frac{\Omega C}{1+\Omega C} \right). \label{eq:freeen}
\end{align}
The first term describes the elastic free energy: $n_p$ is the number of polymer chains per unit dry reference area $A_0$ and $F_{iK}=\frac{\partial x_i}{\partial X_K}$ is the deformation gradient tensor, with $F_{iK} F_{iK}=\mathrm{tr}(\*F^T \*F)=\sum_i \lambda_i^2$ in terms of principal stretches $\lambda_i$. Following standard conventions, deformed/current configurations are denoted by lowercase letters and dry reference configurations are denoted by capital letters unless otherwise noted \cite{goriely}. The second term describes the mixing free energy: $\Omega$ is the area of a solvent molecule, $C$ is the nominal concentration of solvent (number of solvent molecules per unit dry reference area), and $\chi$ is the Flory-Huggins interaction parameter. Both terms are scaled by the product of the Boltzmann constant and temperature, $k_B T$. 

Given that the hydrogel network is held together by permanent cross-links between its polymer chains, we additionally assume that it only changes volume by uptake of solvent, which allows us to express concentrations in terms of the deformation: $\det(\*F)=1+\Omega C$. We require the chemical potential $\mu$ to be zero on the boundary of the hydrogel to mimic submerging it in pure water, as in the experiments. To impose this boundary condition, we perform the standard Legendre transform of Eq. \eqref{eq:freeen} and derive a new free energy $\hat{F}_{\rm{en}}$ in terms of $\mu$ as $\hat{F}_{\rm{en}}(x_i, \mu) =F_{\rm{en}}(x_i, C)-\mu C$ \cite{hong2009inhomogeneous}.  Finally, we model contact by imposing an energy penalty for overlap between the hydrogel and obstacles; as detailed in \textit{Materials and Methods}, our results are insensitive to the choice of this parameter. Note that the chemical potential boundary condition $\mu=0$ is enforced in the regions of contact between the hydrogel and the obstacles. It would be interesting to also consider the impact of dynamic boundary conditions in this problem: for example, a boundary condition requiring no normal solvent flux in the regions where the hydrogel contacts obstacles could be imposed \cite{bouklasfem}. Since we focus on equilibrium quantities in this work, we use a constant chemical potential boundary for simplicity.

\begin{figure*}[htp]
    \centering
    \includegraphics[width=\linewidth]{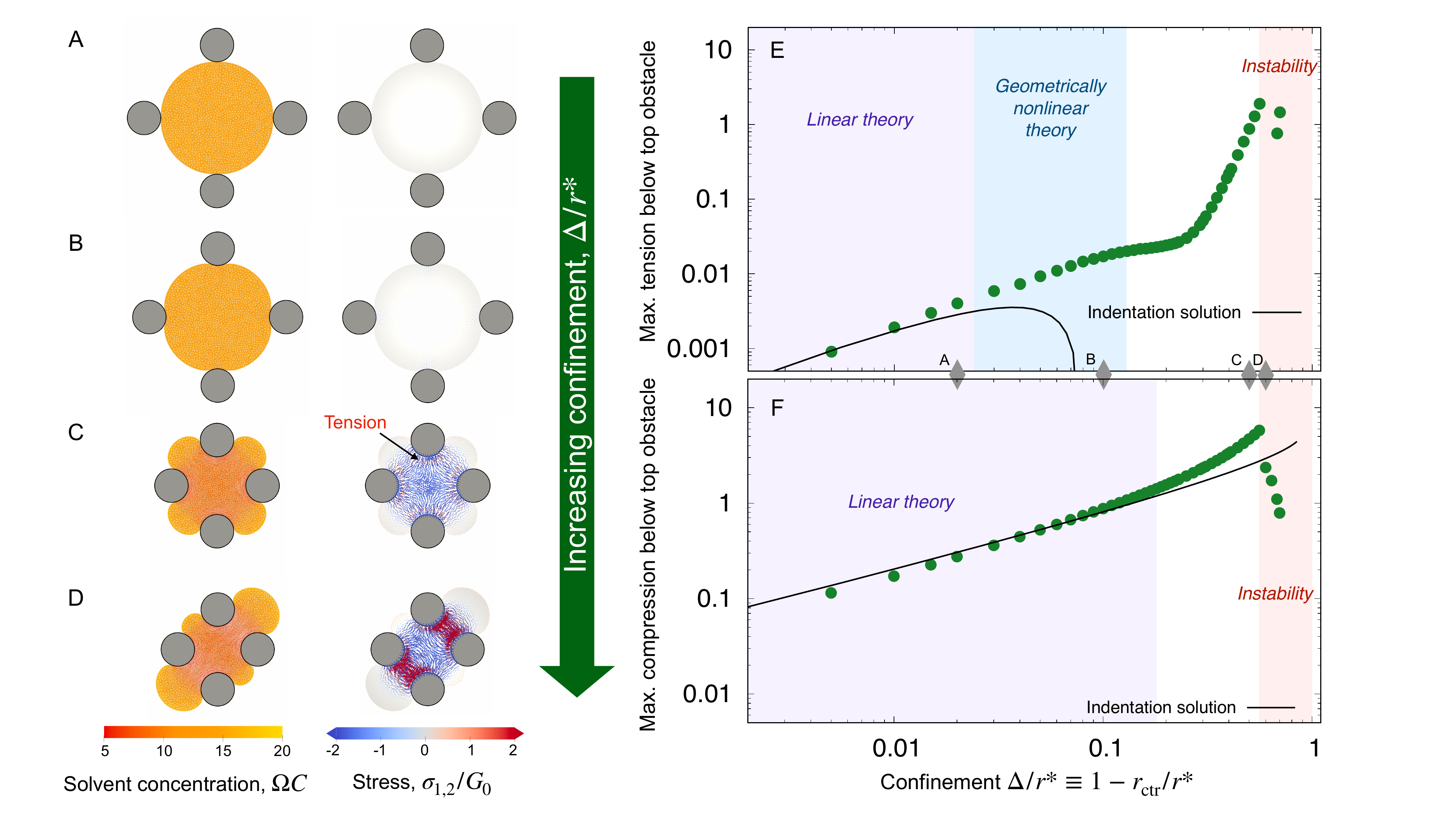}
    \caption{\textbf{Finite element simulations quantify how the equilibrium strains and stresses that develop in the hydrogel depend on obstacle geometry.} (A-D) Maps of the solvent concentration per unit dry reference area with the finite element grid superimposed (left column) and the principal stresses (right column) in the hydrogel at equilibrium, normalized by the area of the solvent molecule $\Omega$ and the fully-swollen hydrogel shear modulus $G_0$, respectively. Confinement increases from top to bottom, in this case by changing the obstacle spacing with fixed $r_{\rm{obs}}/r^* = 0.3$; thus, $\Delta/r^*=1-r_{\rm{ctr}}/r^*=0.02, 0.1, 0.5, 0.6$ from top to bottom, where $\Delta$ is the difference between the unconfined swollen hydrogel radius $r^*$ and the distance to an obstacle from the center $r_{\rm{ctr}}$. Each mesh point in the right column bears two perpendicular lines, oriented and scaled according to the direction and magnitude of the principal stresses $\sigma_1$, $\sigma_2$ at that point, and colored such that compressive stresses are blue and tensile stresses are red. The stresses exceed the range of the color bar for (C-D). (E-F) Points show the maximum principal tensile and compressive stresses obtained from the simulations, taken along the line connecting the hydrogel center to an obstacle center, at this same fixed $r_{\rm{obs}}/r^*= 0.3$ as a function of $\Delta/r^*$. Stresses are again normalized by $G_0$. The values of $\Delta/r^*$ corresponding to (A--D) are marked by the grey diamonds. The predictions of linear indentation theory (solid lines) agree well with the simulation data for small $\Delta/r^*$ (purple shaded region). With increasing $\Delta/r^*$, the maximum tension exceeds linear predictions, but can be reproduced by a \emph{geometrically} nonlinear elastic theory with a linear constitutive law (blue shaded region, SI Sec. H). For compression, the linear theory is accurate over a larger range of  $\Delta/r^*$, but the geometrically nonlinear theory cannot explain the deviations; instead, an elastic model incorporating \emph{material} nonlinearities better captures the data (Eq. \eqref{eq:nh}, SI Fig. S9) . As $\Delta/r^*$ increases even further, the hydrogel exhibits a symmetry-breaking instability (D, red shaded region).  
    }
    \label{fig:sims1}
\end{figure*}

To visualize and track hydrogel deformation during obstructed swelling, along with the concomitant development of internal stresses, we implement this model in FEniCS \cite{alnaes2015fenics}, an open-source finite element platform. Our simulations consider a circular hydrogel swelling around four fixed obstacles. Although our primary focus is the stress distribution in the hydrogel at equilibrium, we simulate the full dynamics of obstructed swelling to provide numerical stability and ensure that there is a realistic path from the initial to the final equilibrium swollen state (SI Sec. E). 

We begin by examining hydrogel swelling around obstacles that have the same size, but varying spacing---just as in the experiments shown in Fig. \ref{fig:schematic}(B,C). Four representative examples, varying from the case of weak to strong confinement, are shown in Fig. \ref{fig:sims1}(A-D). The corresponding maximum tensile and compressive principal stresses on the line connecting the hydrogel center to an obstacle center are plotted in panels (E,F). We quantify these variations using $\Delta/r^*\equiv\left(r^*-r_{\rm{ctr}}\right)/r^*$, where $r^*$ is the equilibrium, fully-swollen, unconfined radius of the hydrogel and $\Delta$ is the difference between $r^*$ and the distance from the center to the closest obstacle $r_{\rm{ctr}}$.  The case of weak confinement in (A) has $\Delta/r^*=0.02$, increasing to $\Delta/r^*=0.6$ for the case of strong confinement in (D). As seen by the color maps of solvent concentration and principal stresses in panel~(A), in this limit of weak confinement, the hydrogel is barely deformed and the resultant stresses are not visible when plotted on the same scale as (B--D). Thus, as we describe in Sec. \ref{sec:weak}, the hydrogel stresses can be captured analytically using linear theories in this regime. Increasing hydrogel confinement to the point shown in panel~(B) causes the magnitudes of the stresses to increase (panels E,F), but they still remain primarily localized near the obstacles; as discussed in Sec. \ref{sec:intermediate1} and Sec. \ref{sec:intermediate2}, the largest compressive stresses can still be described by linear theory, but the largest tensile stresses require consideration of a geometric nonlinearity due to their connection to the rotation of material as it conforms to the obstacle boundary. As the separation between obstacles decreases further, large compressive stresses span the entirety of the hydrogel (panel C), eventually triggering a symmetry-breaking instability (panel D) as discussed in Sec. \ref{sec:strong}.  Finally, while our model is not suitable to treat fracture directly, we discuss the connection between our calculations of stresses and the experimental observations of swelling-induced self-fracture in Sec. \ref{sec:comparison}.

\subsection{Weak confinement}\label{sec:weak}
Consider a hydrogel disk that has swollen around obstacles to reach equilibrium. The chemical potential is spatially uniform and therefore all solvent transport has stopped. Nonetheless, due to contact with the obstacles, the distribution of solvent is inhomogeneous through the hydrogel: solvent preferentially enters the uncompressed lobes of the hydrogel between the obstacles, as shown by the yellow color in Fig.~\ref{fig:schematic}B~\cite{louf_poroelastic_2021}. Now, consider another hydrogel that was first swollen, unobstructed, to its equilibrium size, and then slowly and incrementally squeezed by an identical set of obstacles moving towards its center, acting as four indenters. Solvent must exit the hydrogel where it is indented by the obstacles; recall our condition $1+ \Omega C= \det(\*F)$. For the same final obstacle geometry, these two scenarios must have identical solvent distributions and stress profiles at equilibrium. Thus, the long time limit of obstructed swelling can be treated as an indentation problem, which is well-studied in the limit of small deformations. Making this analogy between obstructed swelling and indentation allows us to apply lessons from a large body of literature on linear contact mechanics and poroelasticity to derive expressions for the stress tensor in the hydrogel \cite{hui2006contact, hu2010using, goriely2016stress, weickenmeier2016mechanics}.

Assuming that deformations relative to the fully swollen state are small, we linearize Eq. \eqref{eq:freeen} to find effective linear elastic parameters in equilibrium (SI Sec. F, \cite{hu2011indentation, bouklasporo}). In particular, for a 2D hydrogel, the effective equilibrium Poisson's ratio $\nu$ and Young's modulus $E$ are given by
\begin{align}
    \nu&=1-2 n_p \Omega\left(n_p \Omega \left(1+\frac{1}{\lambda_0^2}\right) + \frac{1}{\lambda_0^2 (\lambda_0^2-1)}-\frac{2 \chi}{\lambda_0^4}\right)^{-1}\label{poisson},\\
     E&= 2(1+\nu) n_p k_B T \label{youngs},
\end{align}
where $\lambda_0$ is the principal stretch corresponding to the fully swollen state. The expression for the Poisson's ratio can be understood as its value for an incompressible 2D solid, $\nu=1$, minus a correction---reflecting the fact that the compressibility of the swollen hydrogel in equilibrium is generated via solvent transport (i.e., the hydrogel responds to an instantaneous deformation like an incompressible solid before solvent is able to equilibrate). This linearization also yields the shear modulus, $G_0=n_p k_B T$, which we use to normalize stresses throughout this paper. 

\begin{figure}[htp]
    \centering
    \includegraphics[width=\linewidth]{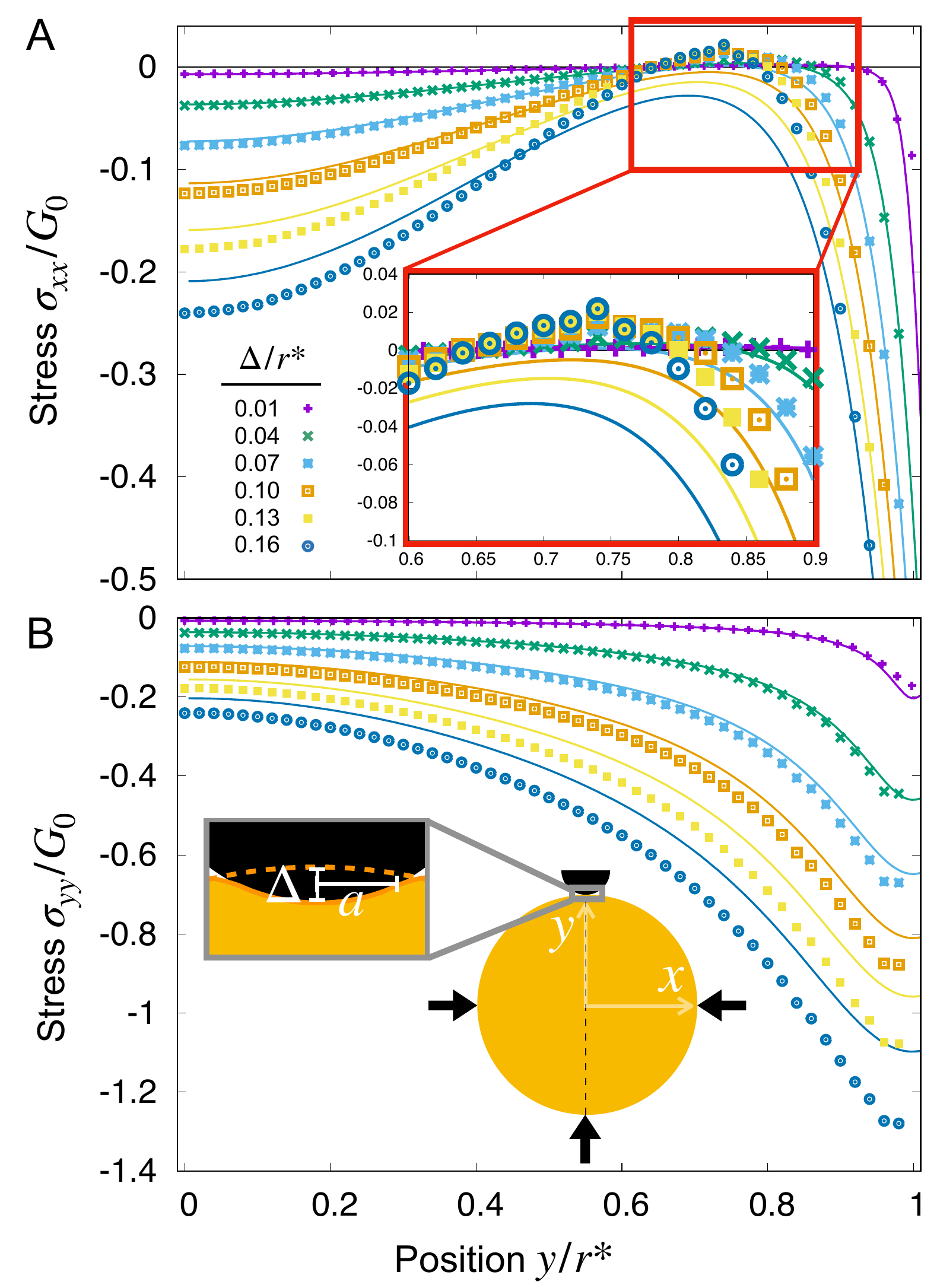}
    \caption{\textbf{When a hydrogel is weakly confined, linear indentation theory can be used to predict the stresses, but misses key features in stronger confinement.} (A, B) Stress components $\sigma_{xx}$ and $\sigma_{yy}$, respectively, determined from simulations (points) and linear theory (lines) along a line running from the center of the hydrogel ($x=y=0$) to the point of contact with the top obstacle ($x=0,~y=r^*$) with fixed $r_{\rm{obs}}/r^*=0.3$. The inset to (B) defines the variables used: the hydrogel-obstacle contact width is $2a$ and the indenter displacement is $\Delta$, defined relative to the undeformed hydrogel radius $r^*$ (dashed orange line). As the indentation depth $\Delta/r^* \equiv 1 - r_{\rm{ctr}} / r^*$ increases, the discrepancy between the linear theory and simulations increases, as expected. With increasing confinement, tension ($\sigma_{xx}>0$) builds up at $r_{\rm{ctr}}/r^* \approx 0.75$, as shown by the magnified inset in A. }
    \label{fig:contactparams}
\end{figure}

Given these effective equilibrium elastic parameters, we solve for the stresses in the hydrogel as a 2D linear contact mechanics problem (SI Sec. G, \cite{barber2018contact,johnson}).
This approach provides expressions for the stress tensor $\sigma_{ij}$ along the line directly beneath the top obstacle as a function of $y$ as shown in Fig. \ref{fig:contactparams}B:

    \begin{align}
    \sigma_{xx}&=\frac{2 \zeta}{\pi} \Bigg(\frac{2 r^{* 3}}{(r^{*2}+y^2)^2} -\frac{1}{r^*}-\frac{2 (r^*-y)}{a^2}\nonumber\\&+ \frac{2 (r^*-y)^2 +a^2}{a^2 \sqrt{(r^*-y)^2 +a^2}} \Bigg),\label{eq:sxx}\\
    \sigma_{yy}&=\frac{2 \zeta}{\pi} \left( \frac{1}{r^*+y} -\frac{1}{r^*}+\frac{2 r^* y^2 }{(r^{*2}+y^2)^2}+ \frac{1}{\sqrt{(r^*-y)^2 + a^2}} \right),\label{eq:syy}\\
    \sigma_{xy}&=0,
    \end{align}
where $\zeta<0$ is the force applied to the indenters and the half contact width $a$ is
\begin{equation}
    a=\sqrt{- \frac{4 \zeta}{E \pi \left( \frac{1}{r_{\rm{obs}}}+\frac{1}{r^*}\right)}},
\end{equation}
as defined in Fig. \ref{fig:contactparams}B. Note that $y$ is defined with respect to the hydrogel's unobstructed, fully-swollen state and ranges from $-r^*$ to $r^*$.

Given these analytical expressions, we next ask: Under what confinement regimes (as parameterized by $\Delta/r^*$) does this linearized theory work, and when does it break down? And since we would like to gain an intuitive understanding of obstructed swelling-induced fracture, how does the maximal principal tensile stress --- which can be used to approximate material strength \cite{gdoutos2020fracture}--- vary with confinement? To address these questions, we first re-cast Eqs.~\eqref{eq:sxx}--\eqref{eq:syy} in terms of $\Delta/r^*$ to facilitate direct comparison with the results of the nonlinear simulations. To do so, we integrate the strain $u_{yy}=\frac{\sigma_{yy}}{Y}-\frac{\nu}{Y} \sigma_{xx}$ using Eqs. \eqref{eq:sxx}--\eqref{eq:syy} to find the displacement at the surface of the indenter relative to the center of the hydrogel, and expand the result to linear order in $a/r^*$. This procedure gives
\begin{equation}
    \Delta = -\frac{\zeta}{E \pi} \left(\ln \left(\frac{16 r^{*2}}{a^2}\right) + \frac{1}{2} (\pi -6 - \pi \nu) \right),
\end{equation}
which we then invert, apply Eqs.~\eqref{poisson}--\eqref{youngs}, and substitute the resulting expression for $\zeta(\Delta)$ into Eqs.~\eqref{eq:sxx}--\eqref{eq:syy}. The resulting expressions for $\sigma_{xx}(\Delta)$ and $\sigma_{yy}(\Delta)$ cannot be expressed in terms of elementary functions, so we omit them, but are shown by the solid lines in Fig.~\ref{fig:contactparams} for the illustrative case of $r_{\rm{obs}}/r^*=0.3$; for comparison, the symbols show the results of the full nonlinear simulations. 

As expected, when $\Delta/r^* \ll 1$, the linearized indentation solution agrees well with the nonlinear simulation results, while as $\Delta/r^*$ increases, the discrepancy between the two becomes more apparent. In particular, as exemplified by the data for $\Delta/r^*=0.16$ (dark blue circles) and $\Delta/r^*=0.13$ (yellow squares) in Fig.~\ref{fig:contactparams}, the linear solution underestimates the compression at both the center $y=0$ and boundary $y=r^*$ of the hydrogel, as well as the tension that builds up at $y/r^* \approx 0.75$ (Fig. \ref{fig:contactparams}A, inset). Indeed, though tension ($\sigma_{xx}>0$) does appear in the linear solutions for small $\Delta/r^*$, it disappears with increasing $\Delta/r^*$, in contrast to the simulation results (see arrow in Fig. \ref{fig:sims1}C, for example). 

Since we are interested in fracture behavior, we thus focus our attention on the maximal value of this tensile stress for the same illustrative case of $r_{\rm{obs}}/r^*=0.3$. By symmetry, we expect the largest stresses to lie beneath each obstacle, since stresses must go to zero at the hydrogel boundary away from obstacles in the weak confinement regime. Moreover, because the straight edges of the finite element mesh can introduce spurious tensile forces at the edge of the hydrogel-obstacle contact (described further in SI Sec. I), we plot the maximum and minimum values of the principal stresses along the $x=0$ line shown in the schematic inset of Fig.~\ref{fig:contactparams}B. The results are displayed in Fig.~\ref{fig:sims1}(E,F). As noted in Fig.~\ref{fig:contactparams}A, the linear theory only predicts the presence of tension for small confinement before deviating from the nonlinear simulation results at $\Delta/r^*\gtrsim0.02$, as shown by the solid line and points in Fig.~\ref{fig:sims1}E, respectively. Interestingly, however, linear theory captures the maximum compressive stress over a broader range of confinement, shown by the solid line in Fig. \ref{fig:sims1}F, which agrees well with the simulation data up to $\Delta/r^* \approx 0.2$. 

Thus, while linear indentation theory can predict both tensile and compressive stresses during obstructed swelling in weak confinement, it underpredicts both for larger deformations---suggesting that the assumptions made in the linear theory are no longer valid. We revisit these assumptions for both tension and compression in the next two sections, respectively. 

\subsection{Tension beyond the linear regime}\label{sec:intermediate1}
The linear theory in the previous section relies on a number of assumptions that can fail as deformations increase:
\begin{enumerate}
\item The effective equilibrium elastic parameters [Eqs. \eqref{poisson}--\eqref{youngs}] are independent of strain,
\item The stress is linearly related to the strain,
\item The strain tensor is linear in the displacements.
\end{enumerate}
To assess the validity of these assumptions, we compare the results of our full nonlinear simulations to those of more complex elastic models that incorporate \emph{material}/\emph{geometric} nonlinearities.

First, we explore the limits of assumption 1 by relaxing assumptions 2 \& 3. Specifically, we compare the hydrogel simulation results to those of a compressible neo-Hookean elastic material with elastic parameters given by Eqs. \eqref{poisson}--\eqref{youngs}. As detailed in SI Sec. F, the neo-Hookean model closely reproduces the stress profiles of the hydrogel simulations over a broad range of $\Delta/r^*$ up to $\approx 0.4$, well beyond the limits of the linear theory at $\sim 0.02$. Therefore, nonlinearities due to the effective elastic parameter mapping can be neglected up to this point. 

Next, we explore the limits of assumption 2 by relaxing assumption 3. Specifically, we use a St. Venant-Kirchhoff elastic model with strain tensor $u_{ij}=\frac{1}{2} \left( \frac{\partial u_i}{\partial x_j} + \frac{\partial u_j}{\partial x_i} + \frac{\partial u_{k}}{\partial x_i} \frac{\partial u_k}{\partial x_j}\right)$; thus, the strain tensor is nonlinear in displacements, but we still require that the hydrogel material follows a linear constitutive law. Note that derivatives are taken with respect to coordinates in the unobstructed, fully-swollen state, denoted here with lowercase letters for simplicity of presentation. Intriguingly, as detailed in SI Sec. H, the St. Venant-Kirchhoff model quantitatively reproduces the maximum principal tensile stress in the hydrogel simulations up to $\Delta/r^*\approx 0.1$, well beyond the limit of the linear theory at $\approx 0.02$. Hence, the excess tension $\sigma_{xx}$ that develops beneath the obstacles just beyond the linear regime is driven by \emph{geometric} nonlinearity, related to the rotations of the hydrogel material as it accommodates the obstacles during swelling, and does not require a nonlinear constitutive relationship. At even larger displacements $\Delta/r^* > 0.1$, the St. Venant-Kirchhoff simulations are unstable and we expect that both geometric and material nonlinearities contribute to the tensile stress. 

How exactly does geometric nonlinearity generate tension during obstructed swelling? We answer this question using an illustrative argument reminiscent of the derivation of the F\"oppl-von K\'arm\'an equation \cite{landau}. As detailed in \emph{Materials and Methods}, we first find the variation of the integrated St. Venant-Kirchhoff strain-energy function with respect to displacements, which can be written in terms of the second Piola-Kirchhoff stress tensor, $\sigma_{ij}^{PK}$. In 2D, this quantity gives the stress component in a material direction $i$ perpendicular to a line that has unit length and normal $j$ in the reference configuration \cite{audoly2010elasticity}. We then make approximations specific to our obstacle geometry. Ultimately, we find
\begin{equation}
    \sigma_{xx}^{PK} \approx -\frac{\partial \sigma_{yy}^{PK}}{\partial y}\left(1-\frac{\Delta}{r^*} \right)r_{\rm{obs}}.\label{eq:sigxx}
\end{equation}
Since $\sigma_{yy}^{PK}$ becomes more negative as $y$ increases, $\frac{\partial \sigma_{yy}^{PK}}{\partial y}<0$ (Fig. \ref{fig:contactparams}B), and therefore $\sigma_{xx}^{PK}>0$ indicating tension. Thus, geometric nonlinearity generates tension beneath an indenter perpendicular to the indentation direction, qualitatively matching our simulations.

\subsection{Compression beyond the linear regime}\label{sec:intermediate2}
A notable result shown in Fig.~\ref{fig:sims1} is that while linear indentation theory predicts tensile stresses for $\Delta/r^*<0.02$, it captures the compressive stresses over a broader range, up to $\Delta/r^*\approx0.2$. Intuitively, this robustness is due to the fact that the large compressive stresses appear perpendicular to the obstacles and are relatively unaffected by rotations of the hydrogel material around the obstacles. Which nonlinearities drive the deviations that arise at even larger displacements? We answer this question by following the same procedure as in the previous section, detailed further in SI Sec. F \& H. In contrast to the case of tension, we 
do not find any parameters for which the St. Venant-Kirchhoff model is more accurate than the linear model. Furthermore, as shown in Fig. S9, past the linear regime, the compressive (Cauchy) stress underneath the top obstacle scales like that of a compressible neo-Hookean elastic material experiencing uniaxial compression, with the principal stretch parallel to indentation set to $\lambda_1=1-\Delta/r^*$ and the principal stretch perpendicular to the indentation set to 1:
\begin{equation}
\sigma_{yy} \sim G_0 \left(1-\frac{\Delta}{r^*}\right) + \frac{2 G_0 \nu}{1-\nu} \frac{\ln\left(1-\frac{\Delta}{r^*}\right)}{\left(1-\frac{\Delta}{r^*}\right)} - \frac{G_0}{\left(1-\frac{\Delta}{r^*}\right)}.\label{eq:nh}
\end{equation}
Thus, unlike the case of tension, deviations from the linear theory do not arise from geometric nonlinearities and can instead be attributed to \emph{material} nonlinearities. 

\subsection{Symmetry-breaking instability}\label{sec:strong}
A striking phenomenon arises in our simulations as the separation between obstacles decreases further: as shown in Fig.~\ref{fig:sims1}D, the hydrogel displays a symmetry-breaking instability and swells preferentially along a diagonal. Why does this instability arise? Inspecting the spatial distribution of compressive stresses provides a clue. As the hydrogel swells in increasing amounts of confinement, its central core becomes increasingly compressed (see, e.g., Fig.~\ref{fig:sims1}C--D). Compressing this circular core along a \emph{single} axis, forming an ellipse, requires less energy than does compressing this core isotropically. Thus, as confinement increases, one expects the case of asymmetric swelling to be energetically preferred, leading to this instability---as described further in SI Sec. J.

\begin{figure}[htp]
    \centering
    \includegraphics[width=\linewidth]{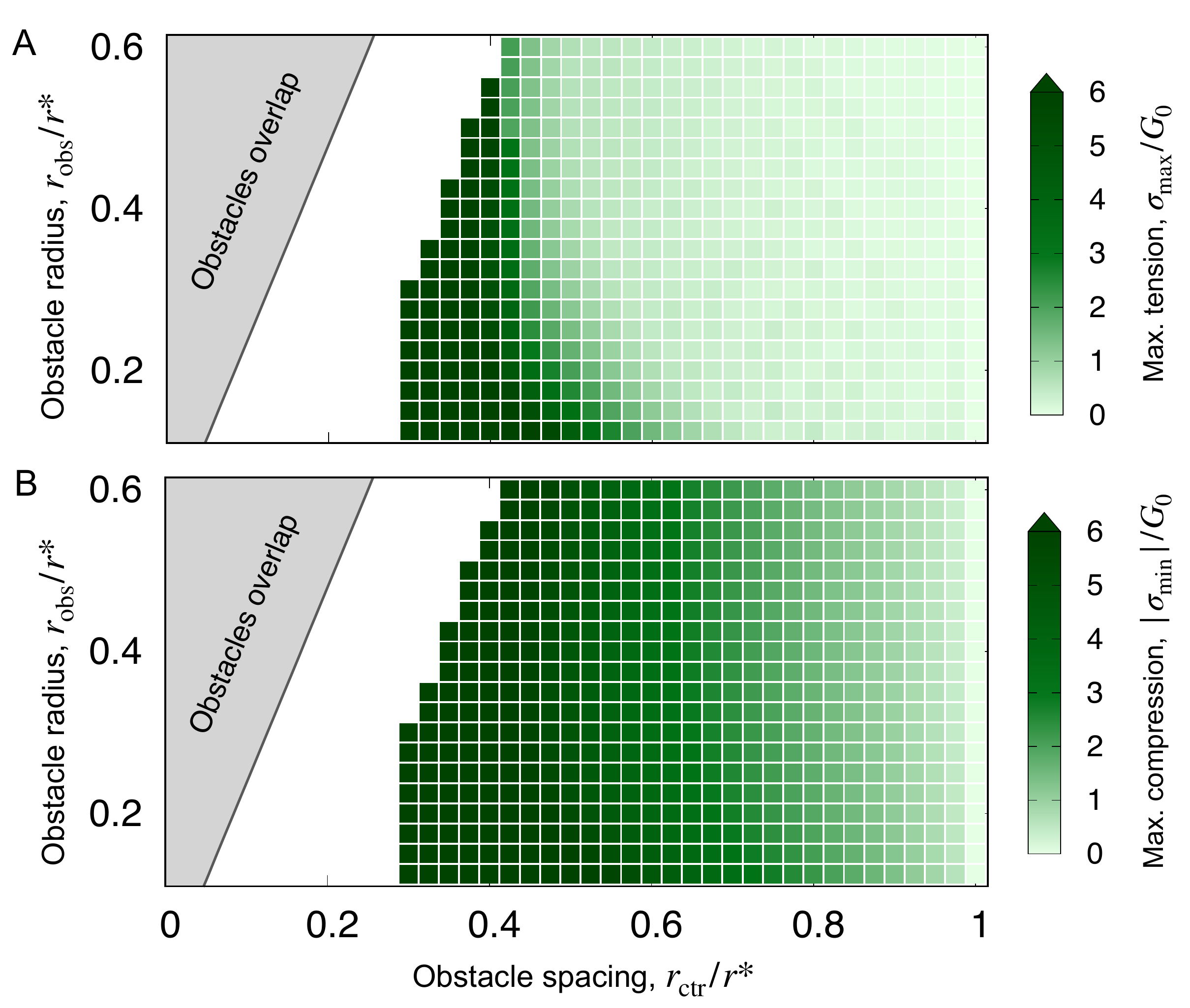}
    \caption{\textbf{Simulations predict that hydrogel stresses grow as obstacles are made smaller and brought closer together in a manner similar to the experimental findings.} (A, B) Most positive (tensile) principal stress $\sigma_{\rm{max}}$ and magnitude of the most negative (compressive) principal stress $|\sigma_{\rm{min}}|$, respectively, again normalized by the fully-swollen shear modulus $G_0$. Each (light or dark) green square corresponds to a simulation. As in Fig. \ref{fig:threshold}, the grey region corresponds to overlapping obstacles. Note that we take the maxima/minima over the entire mesh, rather than just beneath an obstacle, to avoid non-monotonic behavior in the instability regime.}    
    \label{fig:sims4}
\end{figure}

\section{Comparison between simulations and experiments}\label{sec:comparison}
Our theoretical analyses and simulations capture the essential features of the hydrogel deformation observed in experiments (e.g., Figs.~\ref{fig:schematic}B and \ref{fig:sims1}C). They also enabled us to explore how stresses develop during obstructed swelling more generally (Figs.~\ref{fig:sims1}--\ref{fig:contactparams}). While our model is not suitable to directly treat the swelling-induced self-fracture observed experimentally, the simulated stresses help rationalize this phenomenon. To this end, we compare the simulations to the experimental state diagram shown in Fig.~\ref{fig:threshold} by plotting the maximum principal tensile and compressive stresses as a function of $r_{\rm{obs}}$ and $r_{\rm{ctr}}$. The results, shown in Fig.~\ref{fig:sims4}, bear a compelling resemblance to the experimental results. In particular, the convexity and shape of the experimental fracture boundary are similar to the simulated contours of maximum principal stress. Indeed, appealing to a commonly-used fracture criterion for brittle materials \cite{gdoutos2020fracture}, we expect that hydrogel fracture occurs when the maximum tensile stress exceeds a threshold --- whose exact value would establish the position of the experimental fracture boundary in Fig.~\ref{fig:sims4}A. We conjecture that this threshold is reached prior to the symmetry-breaking instability revealed by the simulations, as we do not observe it in our experiments; experiments using tougher hydrogels than those studied here may be a useful way to probe this deformation mode in future work. 

\section{Discussion}
Despite the ubiquity of obstructed growth and expansion in our everyday lives, how exactly this process generates large, spatially non-uniform stresses in a body---and how these stresses influence its subsequent growth/expansion in turn---has remained challenging to systematically study. One reason is the lack of model systems in which this intricate coupling between growth and stress can be probed both experimentally and theoretically. We addressed this need by studying the swelling of spherical hydrogel beads in obstacle arrays of tunable geometries. Our experiments revealed a striking phenomenon: under weak confinement, a hydrogel retains its integrity and assumes a symmetric, four-lobed clover-like shape, while in stronger confinement, it repeatedly tears itself apart as it swells. We elucidated the underlying physics by adopting established models of hydrogel swelling to map the tensile and compressive stresses arising during swelling. In particular, we found that when a hydrogel is weakly deformed, stresses are well-described by linear indentation theory, while as a hydrogel is increasingly deformed, geometric and material nonlinearities engender large tensile and compressive stresses tangential and normal to the obstacles, respectively, driving the hydrogel towards fracture. 

Because our study represents a first step toward fully unravelling the mechanics of obstructed growth and expansion, it necessarily has limitations. For example, while the experimental results shown in Fig.~\ref{fig:threshold} revealed a fracture threshold that varies with obstacle geometry, quantitative comparison to theory and simulation will require more precise control over the system geometry and dimensionality~\cite{joshimorpho, joshi2023energy}, both in experiments and in the model, along with a more detailed treatment of the microscopic processes underlying fracturing~\cite{lai2018probing,kuna2013finite, mao2018theory,wang2012delayed, mao2018theory, baumberger2020environmental, yang2022rate}. Moreover, many more experimental trials near this threshold and with hydrogels of varying mechanical properties, along with high-resolution imaging of crack propagation~\cite{leslie2021gel}, will be useful in characterizing the details of the fracturing process, which likely depend on the presence of microscopic imperfections in the hydrogel. Finally, although here we restricted our attention to the case of rigid obstacles, many scenarios involve growth/expansion around \emph{compliant} constraints---e.g., the development of biofilms, tissues, and organs in the body~\cite{helmlinger1997solid, zhang2021morphogenesis, fortune2022biofilm,goodwin2019smooth}, with potential implications for biological function~\cite{mobius2015obstacles, atis2019microbial}. Extending our study to the case of deformable obstacles would therefore be a useful direction for future work.

Our results may be especially relevant to diverse applications of hydrogels, and other swellable materials, that frequently involve their confinement in tight and tortuous spaces. For example, driven by growing demands for food and water, hydrogels are increasingly being explored as additives to soils to absorb and release water to plants and therefore reduce the burden for irrigation~\cite{misiewicz2022characteristics, woodhouse1991effect, demitri2013potential, guilherme2015superabsorbent, lejcus2018swelling}. They are also widely adopted in other applications, such as oil recovery, construction, mechanobiology, and filtration, all of which involve hydrogel swelling in confinement~\cite{krafcik2017improved, kapur1996hydrodynamic, dolega2017cell, lee2019dispersible}. A common assumption made in all these cases is that the hydrogel retains integrity as it swells; however, our study indicates that these applications should be evaluated for the possibility of swelling-induced self-fracture. Indeed, fracture could lead to the production and dispersal of many small hydrogel fragments, potentially reducing their utility and leading to environmental contamination. This process should therefore be carefully considered in a wide range of real-world contexts.

\section*{Materials and Methods}
\subsubsection*{Experimental design}
To create the obstacle array, we 3D print cylindrical columns in Clear v4 resin using a Form3+ industrial 3D printer (Formlabs), and cut the acrylic plates using an Epilog Laser Mini 24 laser cutter and engraving system.  We secure the columns to the acrylic plates using a twist-and-lock mechanism. The hydrogels are polyacrylamide beads (``water gel beads" obtained from Jangostor) and are stored in a screw cap container prior to experiments; as such, they experience some slight swelling due to ambient humidity. The hydrogel beads have varying sizes and colors, but all appear to have the same swelling behavior and beads of similar sizes are used for experiments (SI Sec. A). These hydrogel beads were extensively characterized in our previous experiments~\cite{louf_under_2021, louf_poroelastic_2021}, which provide additional detail. 

Early in the swelling process (e.g. 2 h in Fig. \ref{fig:schematic}B,C), each hydrogel appears out of focus since it has not yet made contact with the top plate (focal plane for imaging), and cusps are visible on its surface due to differential swelling as water enters the hydrogel from its outer surface~\cite{tanaka1986kinetics, tanaka1987mechanical, bertrand2016dynamics, curatolo2017transient}. We verify that the hydrogel beads swell to an equilibrium shape without rupturing when no obstacles are present (either with or without the 40~mm $z$-confinement), indicating the transient stresses are insufficient to drive fracture as in other less tough gels~\cite{de2016preparation, leslie2021gel}. Because the plates and obstacles are made of acrylic or a polymeric resin, respectively, we do not expect or see evidence of adhesion between the polyacrylamide hydrogel surface and the confining surfaces. For experiments in which we image the entire process of obstructed swelling, we verify that the hydrogel reaches an equilibrium shape (in the cases of no fracturing) when its size/shape does not noticeably change for at least three hours. In the trial shown in Fig.~\ref{fig:schematic}, for example, equilibration took 88~h. In other trials where we do not image the entire process of obstructed swelling, we let the hydrogel swell either until fracture has occurred or for at least 40~h past the time to fracture for a trial with the same $r_{\rm{obs}}$ and a smaller $r_{\rm{ctr}}$. If we already recorded an outcome of `no fracture' for a trial with the same $r_{\rm{obs}}$ and a smaller $r_{\rm{ctr}}$, we waited a minimum of 8 h beyond the equilibration time for an imaged reference trial with similar geometry. Each symbol in Fig. \ref{fig:threshold} represents a single experiment.

\subsubsection*{Simulation design}
We create meshes using FEniCS's built-in mesh generation function with cell size set such that there are 30 vertices along the radius. We confirmed that this mesh resolution was sufficient for numerical convergence. We set $n_p \Omega=0.001$ and $\chi=0.3$. The penalty function is integrated over the hydrogel boundary, and is given by 
\begin{equation}
    \frac{F_{\rm{pen}}}{2 \pi r_0 k_B T}= \frac{p}{4 \pi r_0} \sum_i \langle r_{\rm{obs}}^2-(\*x-\*x_i)^2 \rangle_+^2,
\end{equation}
where $p$ is the penalty strength, $\*x_i$ is position of the center of the $i^{\rm{th}}$ obstacle, $r_0$ is the dry reference radius of the hydrogel, and the sum is over all obstacles. The brackets $\langle \cdots \rangle_+$ take the positive part of the argument, defined as $\langle x \rangle_+= \frac{x + |x|}{2}$. Thus, when evaluated at positions away from any obstacles, the penalty function is zero, but takes a large positive value inside the obstacles. We set $p=6.25 \pi/r_0^4$ to generate the data shown in this text, and have verified that using $p=62.5 \pi/r_0^4$ produces the same results.

Following the suggestions of Refs.~\cite{zhang2009finite, bouklasfem, angbouklas}, we use the backwards Euler scheme for time integration (see SI Sec. E for further discussion of dynamics), Taylor-Hood mixed elements (quadratic elements for the displacement field and linear elements for the chemical potential field), and early time ramping of the chemical potential boundary condition to ensure numerical stability. The Newton-Raphson method is used at each time step, and equilibrium is defined by when zero iterations are required for a step to complete. 

To find the maximum/minimum principal Cauchy stresses, we first calculate the eigenvalues of the stress tensor for a given displacement field. We project these eigenvalue fields onto a function space of discontinuous Lagrange elements of order 1. We then compare the eigenvalues defined on this mesh. The largest positive eigenvalue is the maximum principal (tensile) stress $\sigma_{\rm{max}}$, and the most negative eigenvalue is the minimum principal (compressive) stress, $\sigma_{\rm{min}}$. To find the minimum and maximum stresses beneath an obstacle (data in Fig. \ref{fig:sims1}(E,F)), we instead project the stress tensor onto the vertical line directly beneath the top obstacle. The minimum value of $\sigma_{yy}$ and the maximum value of $\sigma_{xx}$ along this line are plotted as the below top obstacle minimum and maximum stresses respectively.

\subsubsection*{Tension generated by geometric nonlinearity}

We describe the argument leading up to Eq. \eqref{eq:sigxx} in more detail, which demonstrates how tension can appear beneath an obstacle when a nonlinear strain tensor is used. The variation of the integrated St. Venant-Kirchhoff strain-energy function with respect to displacements in terms of the second Piola-Kirchhoff stress tensor is 
\begin{equation}
    \delta W= \int \sigma_{ij}^{PK} \delta u_{ij} dA= \int \sigma_{ij}^{PK}\left(\frac{\partial \delta u_i}{\partial x_j} + \frac{\partial u_k}{\partial x_i}\frac{\partial \delta u_k}{\partial x_j} \right) dA.
\end{equation}
Upon integrating by parts, assuming we cannot vary the displacements at the boundaries due to the presence of obstacles, we find
\begin{align}
   & \delta W= \int \left( \frac{\partial \sigma_{ij}^{PK}}{\partial x_j}\delta u_i +\frac{\partial}{\partial x_j}\left(\sigma_{ij}^{PK} \frac{\partial u_k}{\partial x_i} \right) \delta u_k\right) dA,  \nonumber \\
    &= \int \left( \frac{\partial \sigma_{ij}^{PK}}{\partial x_j}\delta u_i +\frac{\partial \sigma_{ij}^{PK}}{\partial x_j}\frac{\partial u_k}{\partial x_i} \delta u_k + \sigma_{ij}^{PK} \frac{\partial^2 u_k}{\partial x_i x_j} \delta u_k \right)dA. 
\end{align}
Next, in order to make approximations specific to our geometry, we explicitly list all the terms that appear for a two-dimensional solid. Since we will examine stresses directly beneath an obstacle, we set $\sigma_{xy}^{PK}=0$ by symmetry, yielding:
\begin{align}
    \delta W&\approx\int dA \delta u_x \Bigg( \frac{\partial \sigma_{xx}^{PK}}{\partial x}\left(1+\frac{\partial u_x}{\partial x} \right) +\frac{\partial \sigma_{yy}^{PK}}{\partial y}\frac{\partial u_x}{\partial y}\nonumber \\&+\sigma_{xx}^{PK} \frac{\partial^2 u_x}{\partial x^2}+\sigma_{yy}^{PK} \frac{\partial^2 u_x}{\partial y^2} \Bigg)  \nonumber\\
    &+ \int dA \delta u_y \Bigg( \frac{\partial \sigma_{yy}^{PK}}{\partial y}\left(1+\frac{\partial u_y}{\partial y}\right) +\frac{\partial \sigma_{xx}^{PK}}{\partial x}\frac{\partial u_y}{\partial x} \nonumber \\&+\sigma_{xx}^{PK} \frac{\partial^2 u_y}{\partial x^2}+\sigma_{yy}^{PK} \frac{\partial^2 u_y}{\partial y^2} \Bigg). 
\end{align}
In equilibrium, the coefficients of $\delta u_x$ and $\delta u_y$ must be zero (in the absence of body forces). We focus our attention on the coefficient of $\delta u_y$. If we consider the internal stresses directly beneath the obstacle, we can set $\partial u_y/\partial x$ to zero by symmetry. We approximate the deformation as an affine contraction in the $y$ direction, which sets $\partial u_y/\partial y\approx -\Delta/r^*$ and $\partial^2 u_y/\partial y^2\approx 0$. We assume that the curvature of the $y$ displacement, $\frac{\partial^2 u_y}{\partial x^2}$, is determined by the curvature of the obstacle, $1/r_{\rm{obs}}$. With these substitutions, the equilibrium condition becomes
\begin{equation}
     \frac{\partial \sigma_{yy}^{PK}}{\partial y}\left(1-\frac{\Delta}{r^*} \right)+  \frac{\sigma_{xx}^{PK}}{r_{\rm{obs}}} \approx 0.
\end{equation}
Solving for $\sigma_{xx}^{PK}$, we find
\begin{equation}
    \sigma_{xx}^{PK} \approx -\frac{\partial \sigma_{yy}^{PK}}{\partial y}\left(1-\frac{\Delta}{r^*} \right)r_{\rm{obs}}.\label{eq:sigxx2}
\end{equation}
Just beyond the linear regime, $\sigma_{yy}^{PK}$ should follow the same trends as $\sigma_{yy}$ predicted by the linear theory. In Fig. \ref{fig:contactparams}B, we observe that $\sigma_{yy}$ decreases as $y$ increases ($\partial \sigma_{yy}/\partial y<0$) until it plateaus at $y/r^* \approx 1$. Thus, $\partial \sigma_{yy}/\partial y$ reaches its minimum a small distance away from the obstacle boundary, and our argument predicts that the largest geometric nonlinearity-generated tensions will appear there. This location is indeed where the greatest tensile stresses appear in simulations (arrow in Fig. \ref{fig:sims1}C). We can also compare the magnitude of the tension predicted by Eq. \eqref{eq:sigxx2} using $\sigma_{yy}$ from the linear theory with the tension measured in hydrogel simulations--- however, we find an estimate of the tension that is approximately twenty times larger than the maximum tension found via simulations and scales incorrectly with increasing $\Delta/r^*$. These discrepancies are not surprising given the many approximations made in this calculation.

\section*{Conflicts of Interest}
There are no conflicts to declare. 

\section*{Author contributions}: \\
Conceptualization: AP, AK, SSD\\
Investigation: AP, CA, JFL\\
Writing---original draft: AP, AK, SSD\\
Writing---review \& editing: AP, CA, JFL, AK, SSD

\begin{acknowledgments}
It is a pleasure to acknowledge John Kolinski, Amaresh Sahu, and Carolyn Bull for helpful discussions, as well as funding from the NSF through the Princeton MRSEC DMR-2011750, the Project X Innovation Fund, a Princeton Materials Science Postdoctoral Fellowship (AP), and the Camille Dreyfus Teacher-Scholar Program of the Camille and Henry Dreyfus Foundation (SSD). This material is also based upon work by SSD supported by the U.S. Department of Energy’s Office of Energy Efficiency and Renewable Energy (EERE) under the Geothermal Technologies Office (GTO) Innovative Methods to Control Hydraulic Properties of Enhanced Geothermal Systems Award Number DE-EE0009790. \\

\end{acknowledgments}
%
 

\end{document}


\title{Electronic Supplementary Information (ESI)}
\author{}

\maketitle
\vspace{-3em}
\tableofcontents{}
\vspace{-1em}

\subsection{Hydrogel properties}

\subsubsection*{Material properties}
\begin{figure}[htp]
    \centering
    \includegraphics[width=0.65\linewidth]{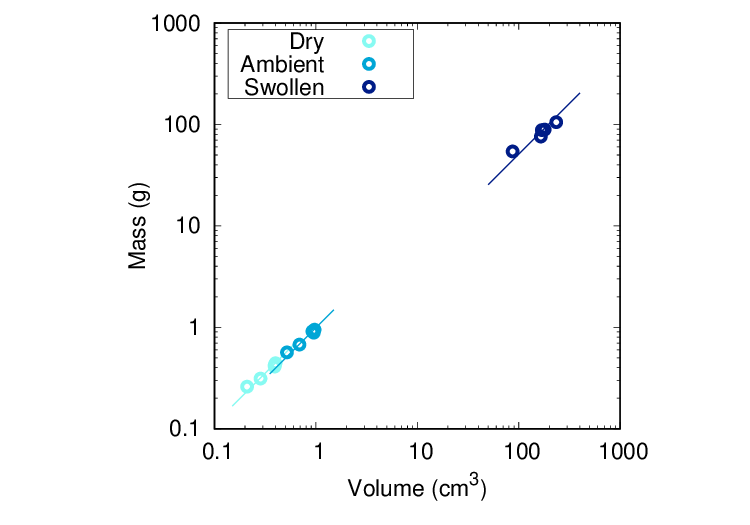}
    \caption{Density is reasonably consistent between different hydrogels in the same state, despite scatter in their mass and volume. The three sets of points correspond to the dry, ambient, and swollen states of the hydrogel. The slope of the line for each set of points gives the average density for each set of five trials. The average densities are 1.12~g/cm$^3$, 0.993~g/cm$^3$, and 0.510~g/cm$^3$ for the dry, ambient, and swollen states respectively.}
    \label{fig:density}
\end{figure}

To assess the consistency of material properties between different hydrogels, we measured the density of five white beads in three different hydration states: ambient, dry, and swollen. The “ambient” state is the condition of the beads prior to any experimentation.  This state is characterized by a small amount of swelling due to ambient humidity, evidenced by slight pliability but high elastic resistance.  The “dry” state was created by placing hydrogel beads in an oven at 100$^{\circ}$C for approximately 24 hours, until the material became completely rigid.  These dry beads were then placed in deionized water for 96 hours to reach their fully “swollen” state. Volumes of the spherical beads for each state were extrapolated from the diameters measured with ImageJ.  The results of these experiments and calculations are displayed in Figure \ref{fig:density}.

\subsubsection*{Size distribution}

\begin{figure}[htp]
    \centering
    \includegraphics[width=0.65\linewidth]{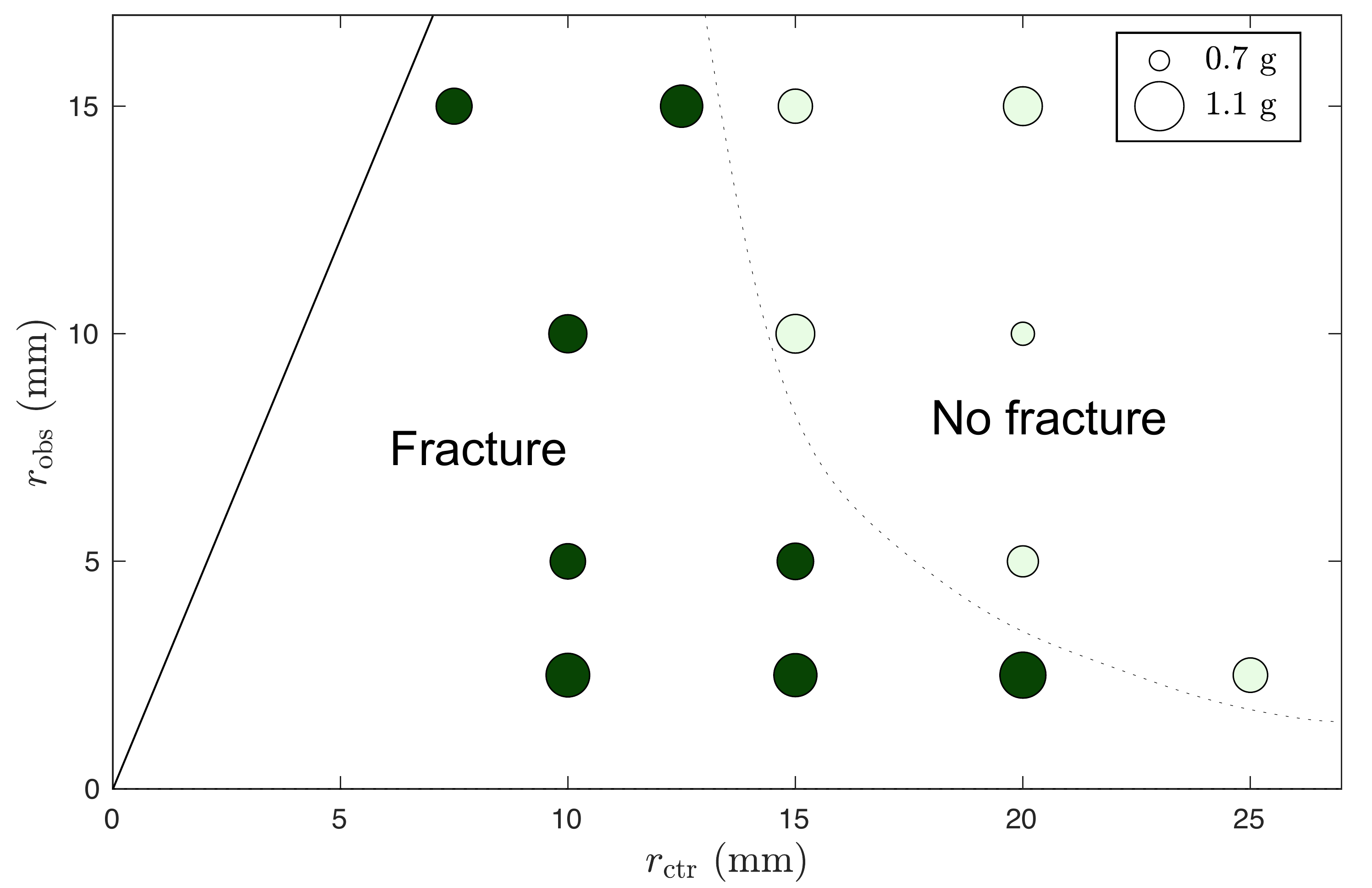}
    \caption{Fracture threshold as a function of obstacle geometry as in Fig. 2 of the main text. The size of the points gives the initial mass of the hydrogel according to the key in the top right, with radius scaled according to mass. The dark green/light green points correspond to trials that fracture/do not fracture respectively, with the dotted line showing the approximate boundary between the two behaviors. The triangular region on the top left shows parameters for which obstacles would overlap.}
    \label{fig:phasesize}
\end{figure}

The hydrogel spheres used in the experiments have slightly different initial masses prior to swelling. This variation in size certainly impacts the location of the reported fracture threshold, as differently sized hydrogels experience different stresses in the same obstacle geometry, and limits our ability to make quantitative statements. In Fig. \ref{fig:phasesize}, we display the fracture data from Fig. 2 of the main text, now scaling each point by the initial mass of the hydrogel. The variation in initial mass between all trials is relatively modest, and the masses of the hydrogels in the trials close to the fracture boundary in particular are similar.

\subsection{Fracture due to vertical confinement}
Hydrogel swelling in the chambers shown in Fig. 1A of the main text is limited both by the presence of obstacles as well as the top and bottom plates. In order to explore the role played by vertical confinement, we allowed hydrogels to swell in chambers with varying $\Delta z$ and no obstacles. 

\begin{figure}[htp]
    \centering
    \includegraphics[width=0.6\textwidth]{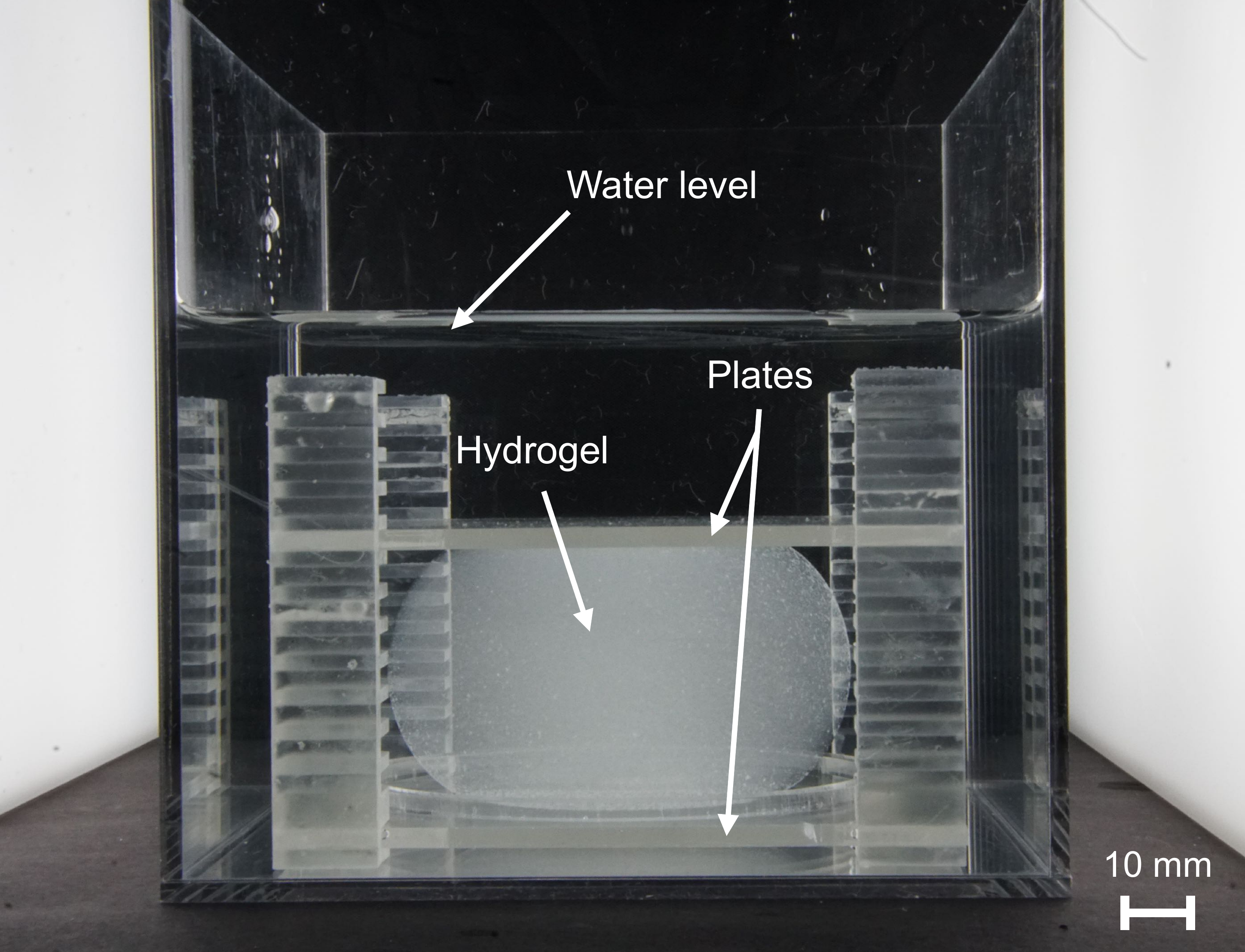}
    \caption{A fully swollen hydrogel remains intact when compressed between two parallel plates spread 40 mm apart.}
    \label{fig: zcon}
\end{figure}

We observed that hydrogels swelling with top and bottom plates separated by $\Delta z< 20$~mm reliably fracture even in the absence of obstacles. In contrast, for $\Delta z=40$~mm, the separation used for all of the experiments presented in this work, we did not observe fracture in the absence of obstacles. One such trial with $\Delta z=40$~mm is shown in Fig.~\ref{fig: zcon}---note that this particular hydrogel had an initial mass of 1.19 g, larger than any of the hydrogels used to create Fig.~2 of the main text.

Understanding fracture as a function of vertical confinement is an interesting direction for future research. Such a study could build on previous theoretical work on flat plate compression of a sphere \cite{tatara1991compression}.

\subsection{3D packing experiments}
\begin{figure}[htp]
    \centering
    \includegraphics[width=0.55\textwidth]{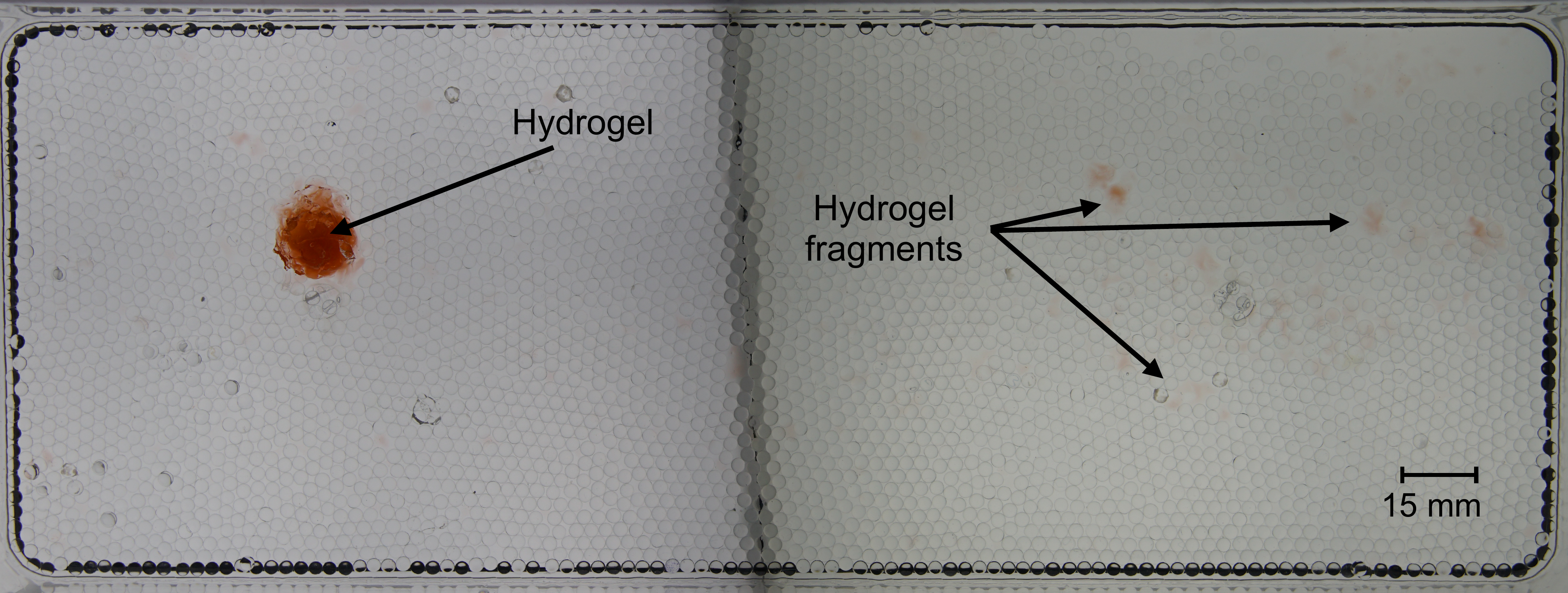}
    \caption{Evidence of rupturing in a hydrogel swollen in a 3D granular medium. On the right side of the image, we see small orange fragments---these are pieces of the hydrogel that have broken off. The fragments are faint because they have expanded significantly relative to the main hydrogel body and are obscured by the glass beads.}
    \label{fig: 3Drupt}
\end{figure}

Following the work of \citet{louf_under_2021}, we allow hydrogel beads to swell in a 3D disordered granular medium composed of borosilicate glass beads of radius 1.5 mm.  As described in the original work, beads of this size mimic coarse, unconsolidated soil.  

We saturate the packing with deionized water and subject the entire column to an applied load of 12 pounds to prevent restructuring of the granular matrix in response to hydrogel swelling.  Since the refractive index of water does not match that of the glass beads, we are unable to collect clear images of the hydrogel during swelling.  However, we can disassemble the packing for visual analysis post-swelling.  

After a short period of 6 hours, we see that pieces of the hydrogel have already broken off and are distributed throughout the packing  (Fig. \ref{fig: 3Drupt}).  When this experiment was replicated with extended swelling periods of 12, 24, and 72 hours, increasing degrees of fragmentation were observed.

Hydrogel swelling can clearly generate stresses large enough to cause rupture in 3D granular media, and hydrogel fragmentation therefore may be relevant in agriculture applications.  Further studies of hydrogels exposed to water in 3D granular media are an interesting and important extension of our current work.

\subsection{Dimensional reduction}\label{sec:2Dsims}
In this section, we provide further discussion on our choice to use a 2D model and discuss the corrections we expect to have due to the 3D nature of our experiments. First, we argue that corrections due to the spherical shape of the hydrogel can be neglected near the midplane and thus locally approximate the spherical hydrogel as a cylinder. Then, we describe how an axially compressed hydrogel cylinder can be modeled in 2D via the generalized plane strain approximation.

\subsubsection*{Neglect of spherical features}

\begin{figure}[h]
    \centering
    \includegraphics[width=0.5\textwidth]{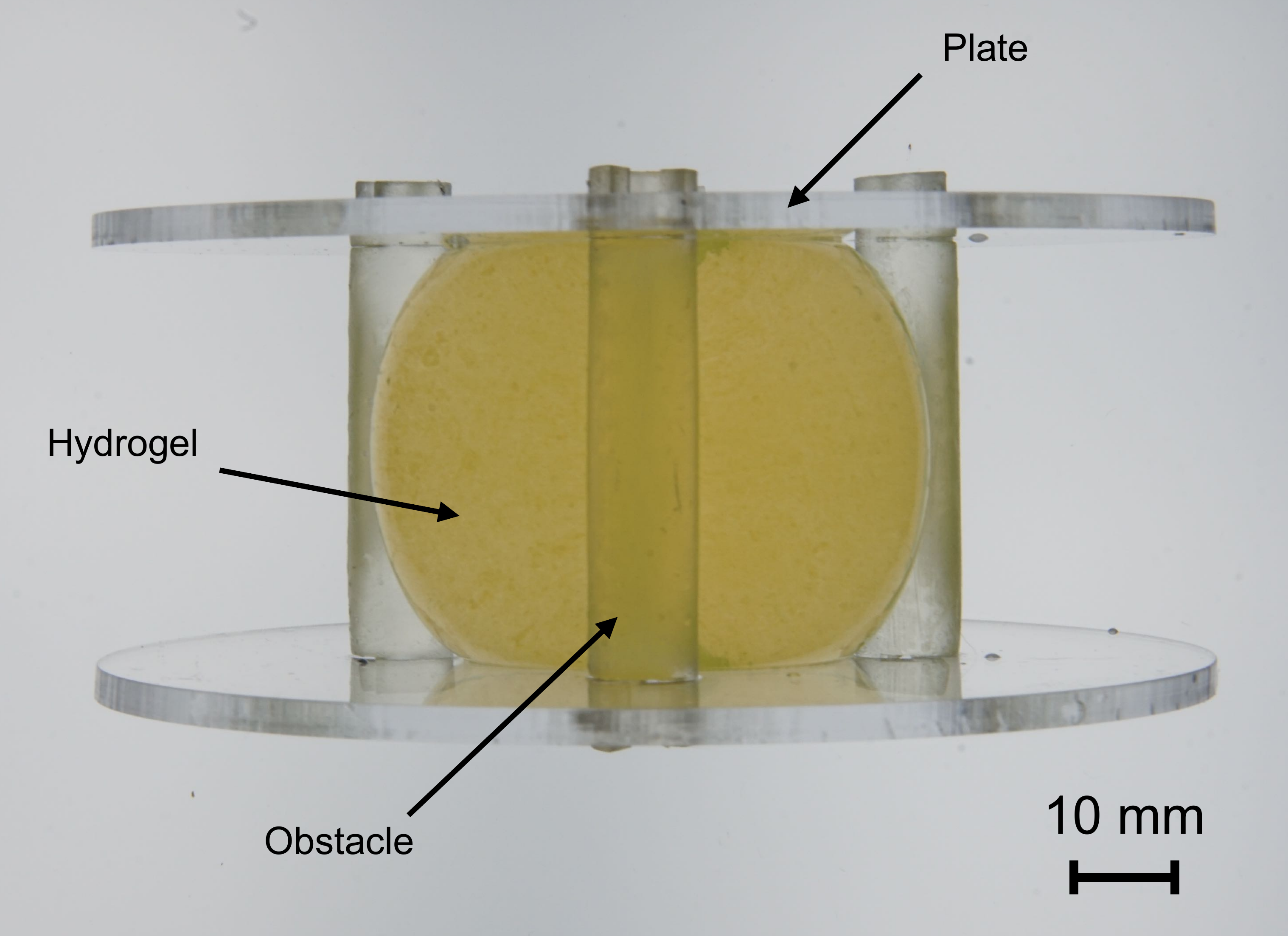}
    \caption{Side view of the hydrogel and obstacles pictured in Fig. 1B of the main text at $\sim90$ hours.}
    \label{fig:sideview}
\end{figure}

To frame our discussion, we refer to Fig. \ref{fig:sideview} which shows the side view of a fully-swollen hydrogel (top view shown in Fig.~1B of the main text). From this viewpoint, we clearly see that the spherical shape preferred by the hydrogel leads to $z$-dependent obstacle-induced stresses---the cylindrical obstacles penetrate further into the hydrogel at the $xy$ plane halfway between the plates compared to the $xy$ plane at the top or bottom plate.

In other words, if we consider horizontal slices of the hydrogel, each slice has a preferred radius that changes as a function of $z$ due to the spherical geometry. We can estimate this preferred radius by neglecting the shape distortion caused by the top and bottom plates (i.e., assuming the hydrogel swells to a perfect sphere in the absence of obstacles). Given a preferred sphere radius $R^*$ and defining $z=0$ to be halfway between the plates, the preferred disk radius $r^*$ of a slice of the hydrogel in the $xy$ plane as a function of $z$ is given by
\begin{equation}
 r^*(z)=\sqrt{R^{*2}-z^2} \approx R^*-\frac{z^2}{2R^*}.
\end{equation}

Indenter displacement is defined as the difference between the preferred hydrogel radius $r^*(z)$ and the leading edge of the obstacle. Since the obstacles are cylinders and their shape/location is independent of $z$, the $z$-dependent preferred hydrogel radius generates a $z$-dependent effective indenter displacement. For a fixed value of $r_{\rm{ctr}}$, the dimensionless indenter displacement as a function of $z$ is
\begin{equation}
    \frac{\Delta(z)}{r^*(z)}=\frac{r^*(z)- r_{\rm{ctr}}}{r^*(z)} \approx \left(1-\frac{r_{\rm{ctr}}}{R^*}\right)- \frac{r_{\rm{ctr}}}{2R^*}\left(\frac{z^2}{R^{*2}}\right).
\end{equation}
When corrections of order $z^2/R^{*2}$ can be neglected, we can ignore the $z$-dependence and locally approximate the hydrogel as a cylinder. This approximation is reasonable close to the midplane of the hydrogel, where the confinement is greatest. 

Thus, we assume our system meets the criteria of a generalized plane strain approximation near the midplane of the hydrogel at equilibrium. We next describe this approximation, first in the context of linear elasticity and then in the context of the nonlinear hydrogel model. We emphasize that this dimensional reduction only holds in equilibrium, as the dynamic features observed in experiments, including initial swelling as well as fracture itself, have 3D features that cannot be straightforwardly described in 2D.  

\subsubsection*{Generalized plane strain: Hookean elasticity}

Generalized plane strain approximations can describe a cylinder in frictionless contact with walls that impose a uniform axial contraction, experiencing tractions independent of $z$ on its sides \cite{saada2013elasticity}. Accordingly, we assume that the strain tensor elements $u_{xx}$, $u_{xy}$ and $u_{yy}$ are independent of $z$, $u_{xz}=u_{yz}=0$, and $u_{zz}=\alpha$, a constant. Assuming Hooke's law holds, we can write down the standard equations for the stress tensor $\sigma_{ij}$ in terms of the strain tensor $u_{ij}$, bulk Young's modulus $E_{3D}$, and bulk Poisson's ratio $\nu_{3D}$ \cite{landau}. 
\begin{align}
    \sigma_{xx}&= \frac{E_{3D}}{(1+\nu_{3D})(1-2\nu_{3D})} \left((1-\nu_{3D}) u_{xx} + \nu_{3D} \left( u_{yy}+ \alpha \right) \right),\\
    \sigma_{xy}&= \frac{E_{3D}}{1+\nu_{3D}} u_{xy}, \\
    \sigma_{yy} &= \frac{E_{3D}}{(1+\nu_{3D})(1-2\nu_{3D})} \left((1-\nu_{3D}) u_{yy} + \nu_{3D} \left( u_{xx}+ \alpha \right) \right), \\
    \sigma_{xz}& =\sigma_{yz}=0,\\
    \sigma_{zz} &=\frac{E_{3D}}{(1+\nu_{3D})(1-2\nu_{3D})} \left((1-\nu_{3D}) \alpha + \nu_{3D} \left( u_{xx}+ u_{yy} \right) \right), 
\end{align}
 Since $\frac{\partial \sigma_{zz}}{\partial z}=\sigma_{xz}=\sigma_{yz}=0$, there are only two equations of equilibrium. Note that $\alpha$ drops out of the equilibrium equations entirely. 

To make comparisons with 2D elasticity, it will also be useful to express the strains in terms of stresses. In order to do this, we write $\sigma_{zz}$ in terms of $\sigma_{xx}$ and $\sigma_{yy}$ using the relation for $u_{zz}$.  
\begin{equation}
    \sigma_{zz}=\nu_{3D} (\sigma_{xx}+\sigma_{yy}) +E_{3D} \alpha. 
\end{equation}
Using this relation, the elements of the strain tensor are 
\begin{align}
    u_{xx}&=\frac{1- \nu_{3D}^2 }{E_{3D}} \sigma_{xx}- \frac{\nu_{3D}(1+\nu_{3D}) }{E_{3D}} \sigma_{yy} - \nu_{3D} \alpha,\\
    u_{xy}&=\frac{(1+\nu_{3D})}{E_{3D}} \sigma_{xy},\\
    u_{yy}&=\frac{1- \nu_{3D}^2 }{E_{3D}} \sigma_{yy}- \frac{\nu_B(1+\nu_{3D}) }{E_{3D}} \sigma_{xx} - \nu_{3D} \alpha,\\
    u_{zz}&=\alpha,\hspace{2 em} u_{yz}=u_{xz}=0.
\end{align}
When $\alpha=0$, we recover the standard plane strain result. 

\subsubsection*{Relation to 2D elastic solid}
We compare the stress and strain tensors derived using the generalized plane strain approximation to those of a 2D elastic solid. For a 2D material, there is no $z$ direction. The two equilibrium equations are therefore identical to those for generalized plane strain. However, the relationship between stress and strain differs. Strain tensor elements are given in terms of stress tensor elements as
\begin{align}
    u_{xx}&=\frac{1}{E_{2D}} \sigma_{xx} - \frac{\nu_{2D}}{E_{2D}} \sigma_{yy},\\
    u_{xy}&=\frac{(1+\nu_{2D})}{E_{2D}} \sigma_{xy},\\
    u_{yy}&=\frac{1}{E_{2D}} \sigma_{yy} - \frac{\nu_{2D}}{E_{2D}} \sigma_{xx},
\end{align}
where $E_{2D}= \frac{4 \mu(\mu+\lambda)}{2\mu+\lambda}$ and $\nu_{2D}=\frac{\lambda}{2 \mu+\lambda}$ are the 2D Young's modulus and Poisson's ratio defined in terms of 2D Lam\'e parameters $\mu$ and $\lambda$. We observe that if we make the substitutions $E_{2D} \rightarrow \frac{E_{3D}}{1- \nu_{3D}^2}$ and $\nu_{2D} \rightarrow \frac{\nu_{3D}}{1-\nu_{3D}}$ and subtract $\nu_{3D} \alpha \delta_{ij}$, we can transform the 2D strain tensor into the generalized plane strain tensor.

 Therefore, if we solve for the stresses using a 2D model, we can create solutions to a corresponding generalized plane strain model using the procedure described above.  

\subsubsection*{Generalized plane strain: nonlinear hydrogel theory}
Thus far, we have only discussed the generalized plane strain approximation in the context of Hookean elasticity.  We show here how the same kinematic assumptions affect the nonlinear hydrogel model when deformations are not necessarily small. 

The free energy density of a 3D hydrogel is \cite{hong2008theory, kang2010variational}:
\begin{equation}
     \frac{F_{\rm{en}}}{V_0 k_B T}=\frac{n_p^{3D}}{2}  \left( F_{iK}^{3D} F_{iK}^{3D} - 3 - 2 \ln(\det(\*F^{3D}))\right)+\frac{1}{v} \left(v C_{3D} \ln \left( \frac{v C_{3D}}{1+v C_{3D}} \right) + \chi \frac{v C_{3D}}{1+v C_{3D}} \right),   
\end{equation}
where $\*F^{3D}$ is the 3D deformation gradient tensor, $v$ is the volume of a solvent molecule, $n_p^{3D}$ is the number of polymer chains per unit reference volume $V_0$, and $C_{3D}$ is the 3D nominal concentration (number of solvent molecules per unit reference volume). 



The Cauchy stress is given by 
\begin{equation}
     \sigma_{ij}=\frac{n_p^{3D} k_B T}{J_{3D}}\left(F^{3D}_{iK}F^{3D}_{jK} -\delta_{ij}\right)+\frac{k_B T }{v J_{3D}}\mathcal{A}(J_{3D})\delta_{ij} - \frac{\mu_{3D} }{v}\delta_{ij},
\end{equation}

where we define the function
\begin{equation}
    \mathcal{A}(J_{3D})\equiv \left(J_{3D} \ln \left( \frac{J_{3D}-1}{J_{3D}}\right) + 1 + \frac{\chi}{J_{3D}} \right).
\end{equation}

We now apply our generalized plane strain assumptions. The deformation gradient tensor and its inverse can be written
\begin{equation}
    \*F^{3D}= \begin{pmatrix} F_{xX} & F_{xY} & 0\\
    F_{yX} & F_{yY} & 0 \\
    0 & 0 & \lambda_z 
    \end{pmatrix}, \hspace{3 em}(\*F^{3D})^{-1}= \begin{pmatrix} F_{yY}/J & -F_{xY}/J & 0\\
    -F_{yX}/J & F_{xX}/J & 0 \\
    0 & 0 & 1/\lambda_z 
    \end{pmatrix},
\end{equation}
where $J$ is the determinant of the 2D deformation gradient tensor $F_{iK}$ and $\lambda_z=1+\alpha$ is the axial stretch. Thus, $J_{3D}=\lambda_z J$. Constrained hydrogel swelling is well-studied---Ref. \cite{zheng2019contact} provides a particularly relevant analysis of indentation on swollen constrained hydrogels.

With these assumptions, $\sigma_{xz}=\sigma_{yz}=0$, and $\sigma_{zz}$ is given by
\begin{equation}
    \sigma_{zz}=\frac{k_B T}{v}\left(\frac{n_p^{3D} v}{\lambda_z J} \left(\lambda_z^2-1 \right) + \frac{\mathcal{A}(\lambda_z J)}{\lambda_z J}  - \frac{\mu_{3D}}{k_B T}\right).
\end{equation}
As in Hookean elasticity, this stress is independent of $z$ and fully determined by the strains in the $x$ and $y$ directions and the prescribed stretch in the $z$ direction. Therefore, $\frac{\partial \sigma_{zz}}{\partial z}=0$ and there are again only two equilibrium equations to solve. In the absence of body forces, these are $\frac{\partial \sigma_{x j}}{\partial x_j}=0$ and $\frac{\partial \sigma_{y j}}{\partial x_j}=0$. 

The other nonzero elements of the Cauchy stress tensor can be written
\begin{equation}
    \frac{\sigma_{ij}}{ k_B T}=\frac{n_p^{3D} F_{iK}F_{jK}}{\lambda_z J}+ \frac{1}{v}\left(\frac{\mathcal{A}(\lambda_z J)}{\lambda_z J} - \frac{n_p^{3D}v }{\lambda_z J} - \frac{\mu_{3D}}{k_B T} \right)\delta_{ij}.\label{3DCauchy}
\end{equation}
In the absence of obstacles, the hydrogel will swell isotropically in $x$ and $y$, reaching an equilibrium swelling stretch ratio $\lambda_0$ that will be a function of $\lambda_z$. 

\begin{figure}[h]
    \centering
    \includegraphics[width=0.55\textwidth]{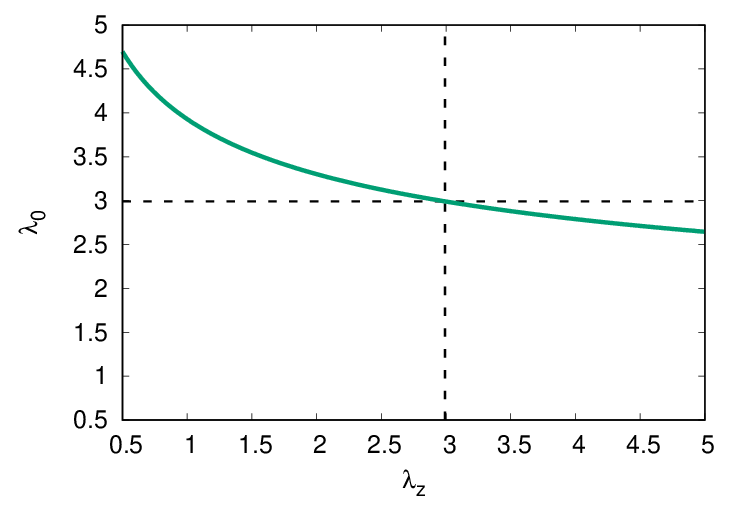}
    \caption{Isotropic transverse stretch $\lambda_0$ as a function of imposed axial stretch $\lambda_z$ for $\mu_{3D}=0, n_p^{3D}v=0.001, \chi=0.3$. The dashed lines show $\lambda_z=\lambda_0$, which corresponds to free swelling in 3D. }
    \label{fig:planestrain}
\end{figure}

We find $\lambda_0$ by setting $\sigma_{ij}=0$ with $F_{iJ}=\lambda_0 \delta_{iJ}$, $J=\lambda_0^2$. This gives the relation
\begin{equation}
    \frac{\mu_{3D}}{k_B T}= \frac{n_p^{3D} v}{\lambda_z}\left(1- \frac{1}{\lambda_0^2} \right)+\frac{1}{\lambda_z \lambda_0^2} + \ln \left( 1- \frac{1}{\lambda_z \lambda_0^2} \right)+\frac{\chi}{\lambda_z^2 \lambda_0^4}.
\end{equation}
This equation must be solved numerically. The relationship between $\lambda_z$ and $\lambda_0$ for $\mu_{3D}=0$, $n_p^{3D} v=0.001$, $\chi=0.3$ is shown in Fig. \ref{fig:planestrain}.
When $\lambda_z=\lambda_0$, we regain the standard 3D result \cite{hong2009inhomogeneous}. 

\subsubsection*{Relation to 2D hydrogel}\label{sec:2dhydrogel}
We compare the stress tensor for a hydrogel in generalized plane strain to that of the purely 2D hydrogel described in the main text. For the energy functional
\begin{equation}
     \frac{F_{\rm{en}}}{A_0 k_B T}=\frac{n_p}{2}  \left( F_{iK} F_{iK} - 2 - 2 \ln(\det(\*F))\right)+\frac{1}{\Omega} \left(\Omega C \ln \left( \frac{\Omega C}{1+\Omega C} \right) + \chi \frac{\Omega C}{1+\Omega C} \right).    
\end{equation}
The corresponding Cauchy stresses are
\begin{align}
    \frac{\sigma_{ij}}{k_B T}=\frac{n_p F_{iK}F_{jK}}{J} + \frac{1}{\Omega} \left(\frac{\mathcal{A}( J)}{J}- \frac{n_p \Omega}{J}- \frac{\mu }{k_B T} \right)\delta_{ij} \label{2D Cauchy},
\end{align}
where
\begin{equation}
    \mathcal{A}(J)=\left( J \ln \left(\frac{J-1}{J} \right)+1 + \frac{\chi}{J} \right).
\end{equation}

We can use 2D solutions to construct generalized plane strain solutions with arbitrary $\lambda_z$. If we make the following substitutions in Eq. \eqref{2D Cauchy}, we regain Eq. \eqref{3DCauchy}.
\begin{align}
    F_{iJ} &\rightarrow \sqrt{\lambda_z} F_{iJ},  \\
    J &\rightarrow \lambda_z J, \\
    n_p &\rightarrow \frac{n_p^{3D}}{\lambda_z},\\
    \Omega &\rightarrow v,\\
    \frac{\mu}{k_B T} & \rightarrow \frac{\mu_{3D}}{k_B T} -\frac{n_p^{3D} v}{\lambda_z J}\left(\frac{1}{\lambda_z}-1\right).
\end{align}
Note that, as in the Hookean elasticity example, these substitutions change the units of various terms---this occurs because 2D and 3D stress have different units.

\subsubsection*{Summary}
Our experimental system can be reasonably modeled in 2D using generalized plane strain assumptions near the hydrogel midsection in equilibrium. In the main text, we work with 2D elastic models to simplify our calculations and simulations. As described above, it is straightforward to transform 2D elasticity solutions to generalized plane strain solutions and vice versa. The imposed stretch in the $z$ direction enters as an effective chemical potential in the hydrogel model, so chemical potential boundary conditions in the 2D simulations would need to be modified to make more quantitative comparisons with experiments.  

\subsection{Dynamics}
We simulate hydrogel swelling dynamically using the kinetic law proposed in \citet{hong2008theory}. 
\begin{equation}
    j_{i}=-\frac{c D}{k_B T} \frac{\partial \mu}{\partial x_{i}},\label{eq:dynamics}
\end{equation}
where $j_i$ and $c$ are the solvent flux and concentration in the current configuration respectively, $\mu$ is the chemical potential, and $D$ is solvent diffusivity. To evolve the concentration field, we discretize the continuity equation using the concentration and solvent flux in the reference configuration: $\frac{\partial C}{\partial t}+\frac{\partial J_K}{\partial X_K} =0$, with $J_K=\det(\*F) \frac{\partial X_K}{\partial x_i} j_i$ \cite{bouklasfem}.

We do not expect the dynamical behavior of our simulations to closely reproduce what we observe in our experiments for several reasons. First, the kinetic law we use is simple and does not account for effects such as changes to the diffusion constant as the polymer network expands. See, e.g., \cite{tanaka1986kinetics, lucantonio2013transient, chester2010coupled}, for further discussions about appropriate expressions for solvent flux. Second, before the hydrogel makes contact with the top and bottom plates, it swells equally in all three dimensions. A 2D simulation therefore only provides a reasonable approximation of equilibrium experimental behavior, as discussed in ESI Sec. \ref{sec:2Dsims}. 

Nevertheless, the simulated 2D dynamics are interesting and complex, and are relevant to hydrogel geometries that remain quasi-2D throughout the swelling process, such as a thin hydrogel disk between two flat plates. We therefore show some characteristic results and provide a brief discussion on simulated dynamical behavior here. In all cases, we initialize the model at a radius $r_0$ and impose an isotropic initial stretch $\lambda_x=\lambda_y=1.5$ to avoid the singularity that appears for the dry state with $\lambda_x=\lambda_y=1$ \cite{hong2009inhomogeneous}. As described in \textit{Materials and methods}, simulations are run until they reach equilibrium which occurs at different times for different trials.

\begin{figure}[htp]
    \centering
    \includegraphics[width=0.65\textwidth]{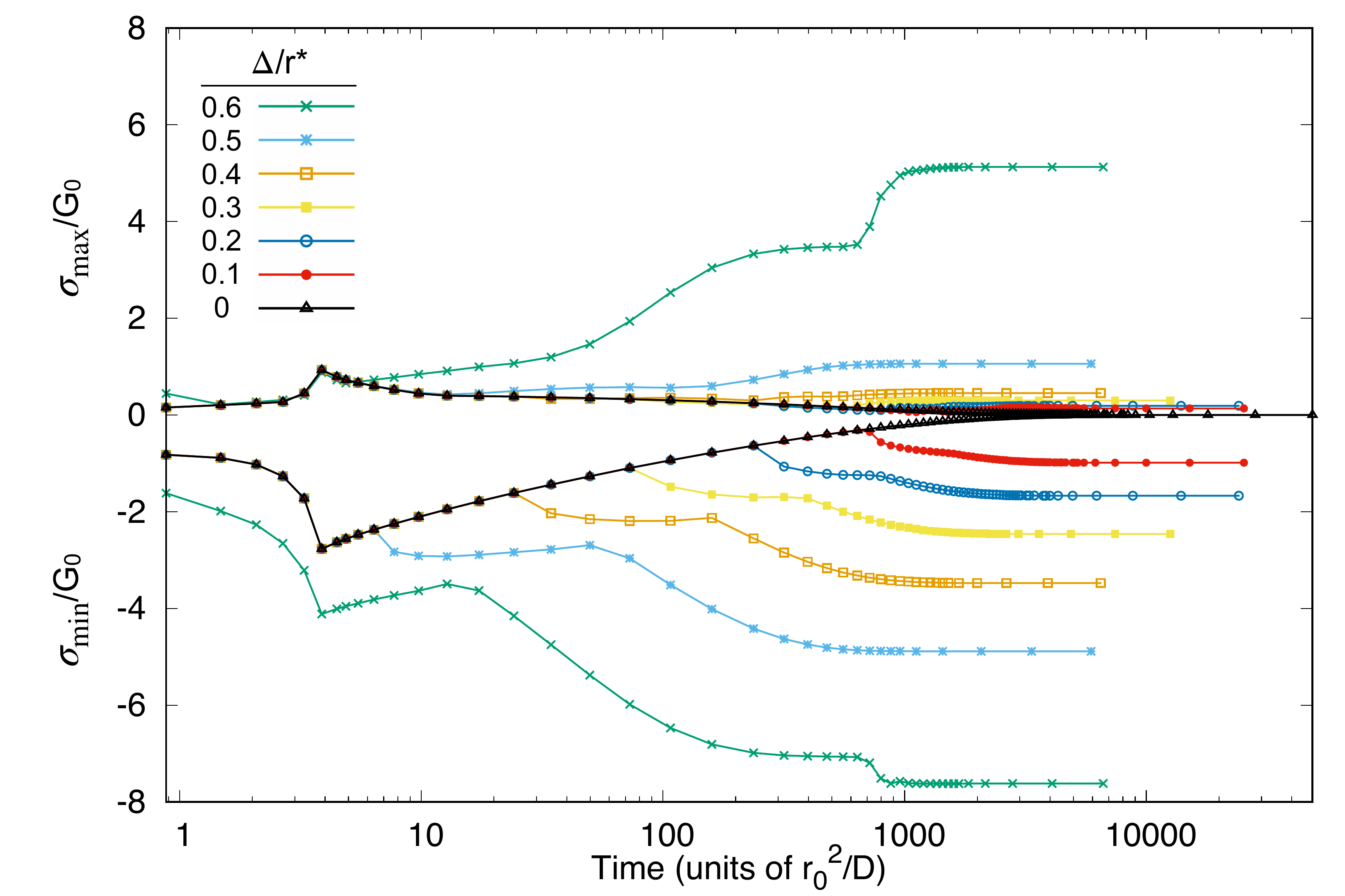}
    \caption{Maximum tensile and compressive principal stresses as a function of time (rescaled by $r_0$, the dry hydrogel radius, and $D$, the diffusivity) for $r_{\rm{obs}}/r^*=0.3$, varying $\Delta/r^*$.}
    \label{fig:dynamics}
\end{figure}

In Fig.~\ref{fig:dynamics}, we show the maximum principal tensile (top set of curves) and compressive (bottom set of curves) stresses as a function of time for different values of $\Delta/r^*$ with $r_{\rm{obs}}/r^*=0.3$. These data are a subset of the data displayed in Fig. 5 of the main text, with maxima/minima taken over the entire mesh. 

To understand these data, we use the $\Delta/r^*=0$ curve as a reference. This curve corresponds to the unconstrained case. We observe a peak in both the maximum compressive and tensile principal stresses at dimensionless time $\hat{t}\equiv \frac{t D}{r_0^2}\approx 4$, followed by a decay to zero. This peak in stress occurs because the outer regions of the hydrogel swell faster than the inner regions, and can lead to the formation of transient cusp structures, as described in the main text and elsewhere \cite{curatolo2017transient}. For $\Delta/r^*\leq 0.5$, simulations follow the free swelling curve until they make contact with the obstacles. Following contact, the hydrogel develops stresses that do not disappear at equilibrium. The data for $\Delta/r^*=0.6$, the strongest confinement shown here, has a number of interesting differences from the other trials. This hydrogel encounters obstacles prior to the time points shown in Fig. \ref{fig:dynamics}, and thus sustains larger stresses earlier compared to the other trials. At $\hat{t} \approx 700$, we see a sharp increase in both the compressive and tensile stresses in this trial, not observed in other trials---this is the symmetry-breaking instability discussed in Sec. IIID of the main text. 

\begin{figure}[htp]
    \centering
    \includegraphics[width=0.65\textwidth]{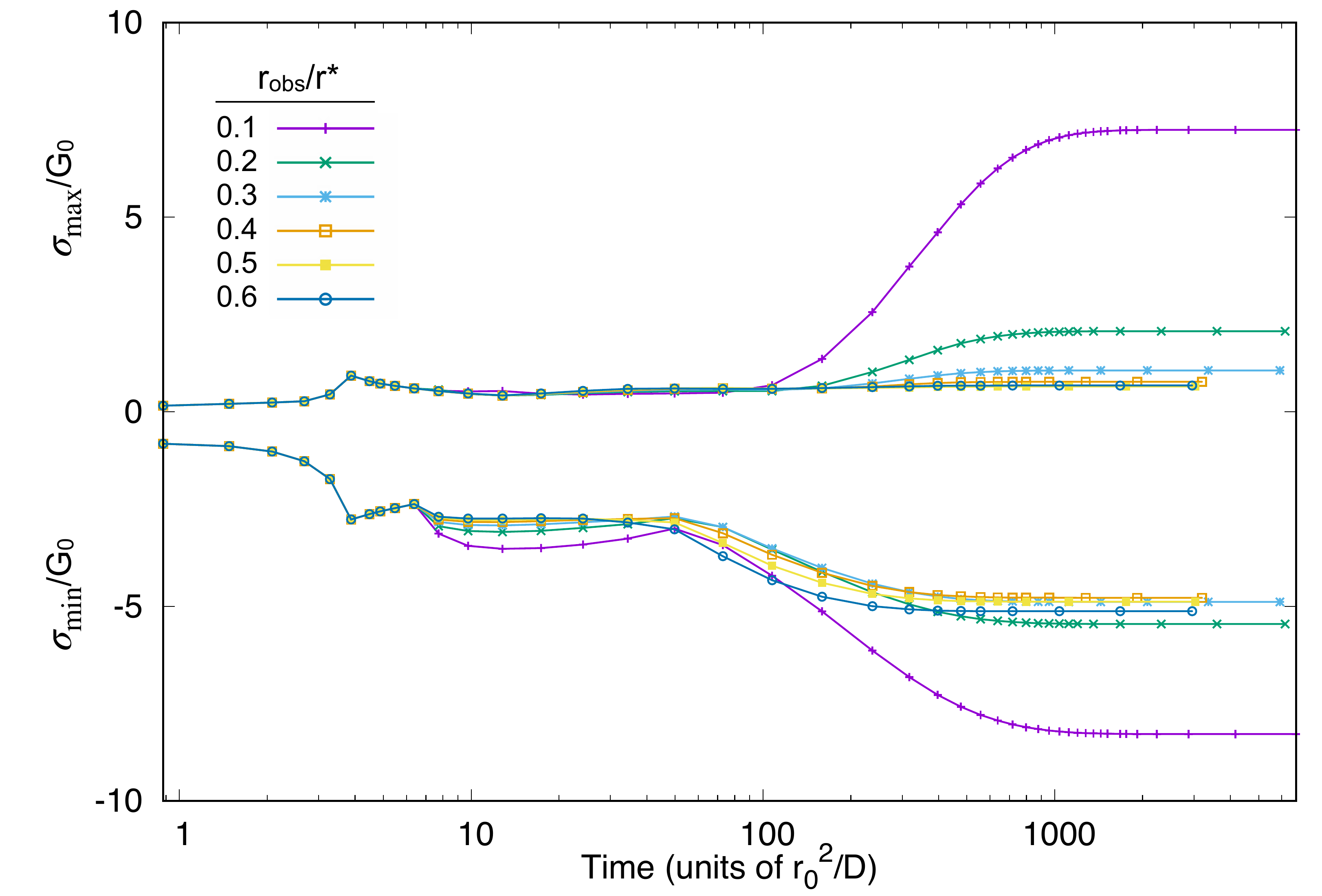}
    \caption{Maximum tensile and compressive principal stresses as a function of time (rescaled by $r_0$, the dry hydrogel radius, and $D$, the diffusivity) for $\Delta/r^*=0.5$, varying $r_{\rm{obs}}/r^*$. The light blue curves in this figure and Fig. \ref{fig:dynamics} display the same data.}
    \label{fig:dynamics2}
\end{figure}

We compare these dynamics to those in Fig. \ref{fig:dynamics2}. Here, we show principal stresses as a function of time with $\Delta/r^*$ held fixed 
 and $r_{\rm{obs}}/r^*$ varied, again with maxima/minima taken over the entire mesh. These data correspond to vertical slices of Figs. 5(A,B). Since $\Delta/r^*$ is the same for all trials, the hydrogels all encounter obstacles at the same time, $\hat{t} \approx 6$. After that time, trials behave differently due to the different curvature of the obstacles.

\subsection{Elastic moduli}\label{sec:elasticmoduli}
In this section, we provide the derivation for the effective Poisson's ratio and Young's modulus of a 2D hydrogel in equilibrium (Eqs. (2) and (3) of the main text) by adapting the argument presented in \citet{bouklasporo}. We begin with the Cauchy stress for a 2D hydrogel model,
\begin{equation}
        \frac{\Omega \sigma_{ij}}{k_B T}=\frac{n_p \Omega}{J} F_{iK}F_{jK}+ \left(\frac{\mathcal{A}( J)}{J}- \frac{n_p \Omega}{J}- \frac{\mu }{k_B T} \right)\delta_{ij},
\end{equation}
with
\begin{equation}
    \mathcal{A}(J)=\left( J \ln \left(\frac{J-1}{J} \right)+1 + \frac{\chi}{J} \right).
\end{equation}
We apply a uniaxial stress in the $x$ direction and require $\sigma_{yy}=0$. This sets
\begin{align}
         0=\frac{\Omega \sigma_{yy}}{k_B T}&=\frac{n_p \Omega}{\lambda_x \lambda_y} \left(\lambda_y^2 -1 \right)+ \left(\frac{\mathcal{A}( \lambda_x \lambda_y)}{\lambda_x \lambda_y}- \frac{\mu }{k_B T} \right) . \label{eq:sigyy0}
\end{align}
We assume that the resulting stretches are small deformations relative to the stress-free equilibrium state with stretch $\lambda_0$, defined for the 2D hydrogel via the equation
\begin{equation}
        \frac{\mu} {k_B T}=\frac{n_p \Omega}{\lambda_0^2}\left(\lambda_0^2- 1 \right) + \frac{\mathcal{A}(\lambda_0^2)}{\lambda_0^2}= n_p \Omega\left(1- \frac{1}{\lambda_0^2} \right) + \ln \left( 1- \frac{1}{\lambda_0^2} \right)+\frac{1}{\lambda_0^2}+\frac{\chi}{\lambda_0^4} \label{eq:l02D}.
\end{equation}
This allows us to define the strain tensor elements in terms of stretches as
\begin{align}
    \lambda_x&=\lambda_0(1+u_{xx}),\\
    \lambda_y&=\lambda_0(1+u_{yy}).
\end{align}
We next expand Eq. \eqref{eq:sigyy0} to linear order in $u_{xx}$ and $u_{yy}$.After simplifying using Eq. \eqref{eq:l02D}, our result is
\begin{equation}
u_{xx} \left( n_p \Omega \left( 1- \frac{1}{\lambda_0^2} \right) + \frac{2 \chi}{\lambda_0^4} - \frac{1}{\lambda_0^2 (\lambda_0^2-1)} \right)=u_{yy} \left( n_p \Omega \left(1+\frac{1}{\lambda_0^2}\right)+ \frac{1}{\lambda_0^2 (\lambda_0^2-1)} - \frac{2 \chi}{\lambda_0^4} \right).
\end{equation}
Defining the Poisson's ratio as
\begin{equation}
    \nu= -\frac{u_{yy}}{u_{xx}},
\end{equation}
we find
\begin{equation}
    \nu= \frac{n_p \Omega \left( \frac{1}{\lambda_0^2} -1 \right) + \frac{1}{\lambda_0^2 (\lambda_0^2-1)}- \frac{2 \chi}{\lambda_0^4} }{ n_p \Omega \left(\frac{1}{\lambda_0^2}+1 \right) + \frac{1}{\lambda_0^2 (\lambda_0^2-1)}- \frac{2 \chi}{\lambda_0^4} }=1-\frac{2 n_p \Omega}{n_p \Omega \left(1+\frac{1}{\lambda_0^2}\right) + \frac{1}{\lambda_0^2 (\lambda_0^2-1)}-\frac{2 \chi}{\lambda_0^4}}.\label{eq:nueff}
\end{equation}

To find the Young's modulus, we perform the same operations in the equation for $\sigma_{xx}$:
\begin{equation}
    \frac{\Omega \sigma_{xx}}{k_B T}=\frac{n_p \Omega}{\lambda_x \lambda_y}\left( \lambda_x^2 -1 \right)+ \left(\frac{\mathcal{A}( \lambda_x \lambda_y)}{\lambda_x \lambda_y}- \frac{\mu }{k_B T} \right).
\end{equation}
We substitute $u_{yy}=-\nu u_{xx}$ and expand in $u_{xx}$ to linear order. After simplifying, we find
\begin{equation}
    \frac{\Omega \sigma_{xx}}{k_B T}=2(1+\nu) n_p \Omega u_{xx}.
\end{equation}
Thus, the effective 2D Young's modulus is
\begin{equation}
    E=2 (1+\nu)n_p k_B T , \label{eq:yeff}
\end{equation}
and the effective 2D shear modulus is
\begin{equation}
    G_0\equiv \frac{E}{2 (1+\nu)}=n_p k_B T.  \label{eq:shear}
\end{equation}

\subsubsection*{Robustness of mapping}
In the derivation of effective elastic parameters, we neglect terms quadratic in $u_{ij}$. Thus, for large deformations, we expect our mapping between hydrogel model parameters and linear elastic parameters [Eqs. \eqref{eq:nueff} and \eqref{eq:shear}] to become inaccurate (i.e., the effective Poisson's ratio and Young's modulus will acquire a dependence on strain). This additional source of nonlinearity has the potential to complicate our comparisons between the nonlinear hydrogel model and the St. Venant-Kirchhoff model in Sec. IIIB and C of the main text. For example, an inhomogeneous strain state could induce spatially-varying effective elastic parameters. However, using simulations of a neo-Hookean elastic model, we argue that this nonlinearity can be neglected for the deformations considered in this work. 

We simulate a compressible neo-Hookean elastic model in FEniCS, using the same mesh resolution as in the hydrogel trials (approximately 30 vertices across the radius; see
Fig. 3, left column). The strain-energy density function is
\begin{equation}
    W=\frac{G_0}{2}\left( F_{iK} F_{iK} - 2 - 2 \ln(\det(\*F))\right)+\frac{G_0 \nu}{1-\nu} \ln(\det (\*F))^2, \label{eq:compneohook}
\end{equation}
where the 2D shear modulus $G_0$ and Poisson's ratio $\nu$ are set according to Eqs. \eqref{eq:nueff} and \eqref{eq:shear}. Setting the coefficients in this manner ensures consistency with linear elasticity. Displacements are defined relative to the zero stress reference configuration with radius $r^*$. 

\begin{figure}[htp]
    \begin{center}
\includegraphics[width=\textwidth]{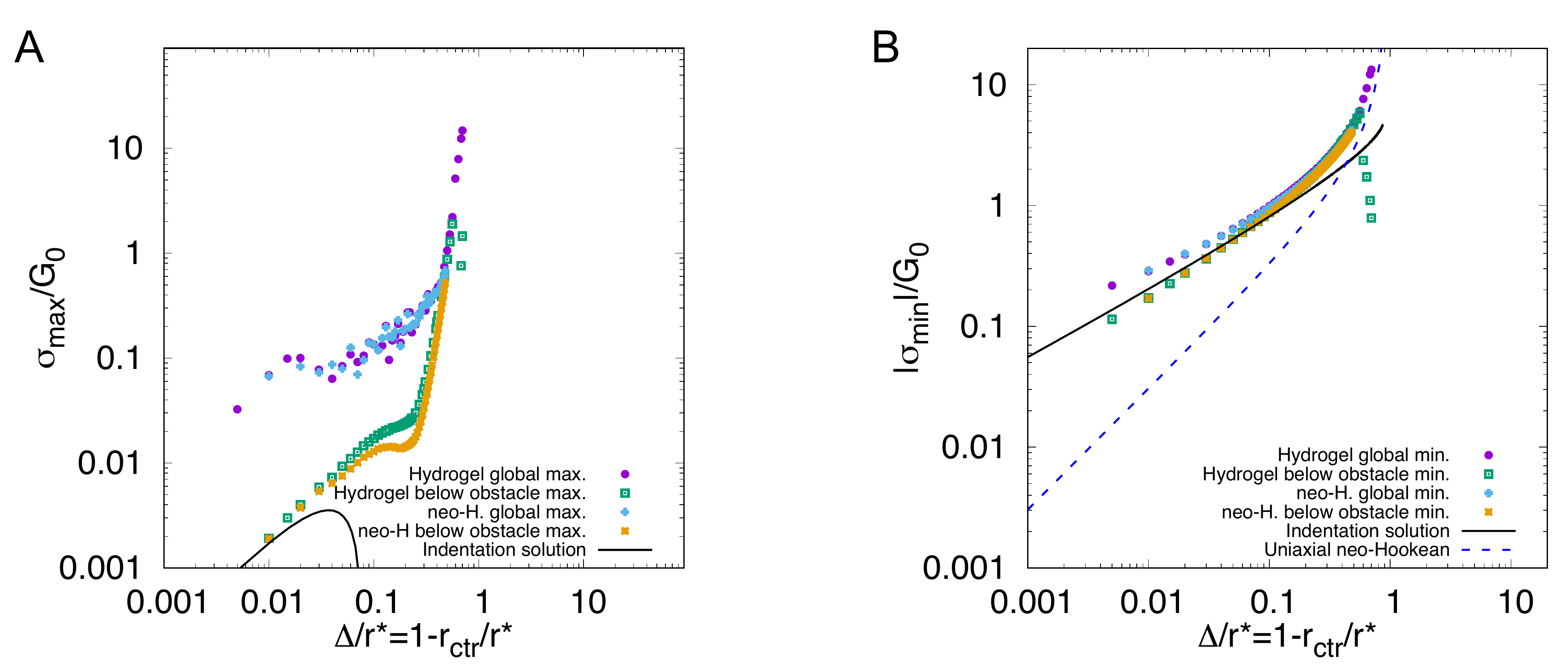}
\end{center}
    \caption{Comparison between maximum principal Cauchy tensile and compressive stresses in simulations with the nonlinear hydrogel model and neo-Hookean elastic model. Results for the neo-Hookean model are shown through $\Delta/r^*=0.48$. Even for these large deformations, we observe reasonably good agreement between the two models. Note that we plot both the maximum/minimum principal stress over the entire hydrogel domain (labeled global max./min.) as well as the maximum/minimum principal stress along the line connecting the hydrogel center to an obstacle center (labeled below obstacle max./min.). For further discussion of the differences between these two measures, see ESI Sec. \ref{sec:maxmax}. In panel B, the dashed line shows Eq. 10 of the main text for comparison.}
    \label{fig:nHmapping}
\end{figure}

\begin{figure}[htp]
    \begin{center}
        \includegraphics[width=\textwidth]{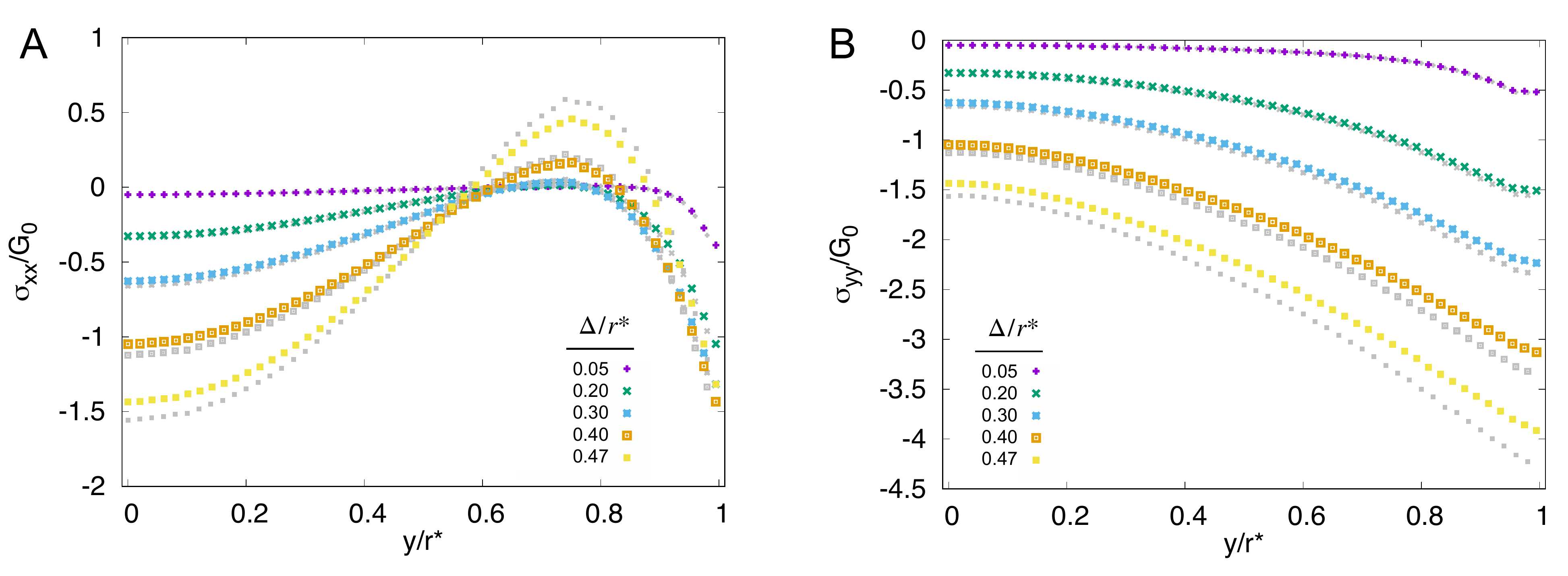}
\end{center}
    \caption{Comparison between Cauchy stress profiles in simulations with the nonlinear hydrogel model and neo-Hookean elastic model, plotted against undeformed fully-swollen coordinates (corresponding to an unobstructed equilibrium state for the hydrogel). Neo-Hookean results are shown with large colored markers, while hydrogel results are shown in grey. Data corresponding to the same value of $\Delta/r^*$ have the same marker shape. Although we clearly see differences between the two models at large $\Delta/r^*$, the two models produce similar stress profiles.}
    \label{fig:nHprofiles}
\end{figure}

In Fig. \ref{fig:nHmapping}, we compare the maximum principal stresses in the neo-Hookean elastic model and the hydrogel model. Although there are some differences between the two data sets, it is difficult to tell whether these differences increase as a function of strain as expected. We learn more by comparing the stress profiles directly in Fig. \ref{fig:nHprofiles}. As $\Delta/r^*$ increases, we clearly see differences between the models increase. However, deviations only appear at the largest values of $\Delta/r^*$ tested, and even then, they remain small relative to the magnitude of the stress. These comparisons suggest that our mapping from hydrogel parameters to elastic parameters is reasonably accurate, perhaps even surprisingly so, for large indentation-type deformations.

\subsection{Indentation solutions}
Here we derive the indentation theory equations in Sec. IIIA of the main text. We start with the Flamant solution applied to a point force $\zeta$ acting normal to an elastic half space in 2D \cite{barber}. We assume the half space fills the region $y<0$ and use the convention that $\zeta<0$ corresponds to a force that compresses the plane. In radial coordinates with the origin at the location at which the point force acts, this is a simple radial stress distribution
\begin{equation}
    \sigma_{rr}=-\frac{\zeta \sin \theta}{\pi r}.
\end{equation}
Directly beneath the point source, $\theta=-\pi/2$ and $\sigma_{rr}=\frac{\zeta}{\pi r}<0$. 

The displacements due to a point force grow logarithmically in 2D. Therefore, we must be mindful of boundaries, as the finite size of our deformable material will influence our calculations. We can gain an appreciation for this subtlety by considering two diametrically opposed point forces acting on the top and bottom of a 2D elastic disk, as in \citet{timoshenko} p 107. In this case, all points on the boundary of the disk experience isotropic compression due to the pair of point forces, and a uniform tension $\sigma_{ij}=-\frac{\zeta}{\pi r^*}\delta_{ij}$ must be added to maintain a stress-free boundary.  By the same argument, $\sigma_{ij}=-\frac{2 \zeta}{\pi r^*}\delta_{ij}$ must be added to the stress tensor to free the boundaries when four perpendicular point forces are acting on the elastic disk. 


We now consider the specific geometry we are interested in: four circular indenters of radius $r_{\rm{obs}}$ acting on an elastic disk of radius $r^*$. In the weak confinement regime, we expect the greatest stresses to develop on the line connecting the center of the elastic disk and the center of an indenter, as stresses go to zero at the points at which the disk and indenter lose contact. Therefore, by symmetry we only need to solve for stresses along this line ($x=0$ line in inset to Fig.~4B in the main text). Appealing to Saint-Venant's principle, we simplify our calculation by modeling one obstacle as a 2D Hertzian indenter and the other three obstacles as point forces. We proceed by generalizing the calculation in \citet{johnson} p129. 

We place the origin at the center of the contact between the elastic body and the Hertzian indenter, and let $b>0$ be the distance from this origin along the centerline with $x=0$. The stress tensor contributions from the point indenters along the line of interest are
\begin{align}
    \sigma_{xx}&= \frac{4 \zeta r^{*3}}{\pi (r^{*2} +(r^*-b)^2)^2},\\
    \sigma_{yy}&= \frac{2 \zeta}{\pi(2 r^*-b)}+\frac{4 \zeta r^*(r^*-b)^2}{\pi(r^{*2}+(r^*-b)^2)^2},\\
    \sigma_{xy}&=0.
\end{align}
The Hertzian indenter contributes
\begin{align}
    \sigma_{xx}&= \frac{\zeta}{\pi} \left( \frac{2 (a^2+2 b^2)}{a^2\sqrt{a^2+b^2}}- \frac{4 b}{a^2} \right),\\
    \sigma_{yy}&= \frac{2 \zeta}{\pi \sqrt{a^2+b^2}},\\
    \sigma_{xy}&=0,
\end{align}
where the contact length $2a$ is given according to
\begin{equation}
    a^2=-\frac{4 \zeta}{E \pi} \left( \frac{1}{r_{\rm{obs}}}+\frac{1}{r^*}\right)^{-1}.
\end{equation}

The stress tensor is the sum of these contributions, plus the isotropic tension from the corrective solution, $\sigma_{ij}=-\frac{2 \zeta}{\pi r^*} \delta_{ij}$. With the substitution $y=r^*-b$, we find Eqs. (4)-(7) of the main text.

The strain $u_{yy}$ can be found as
\begin{align}
    u_{yy}&=\frac{\sigma_{yy}}{E}-\frac{\nu}{E} \sigma_{xx}.
\end{align}
We find the displacement at the surface directly beneath the indenter by integrating
\begin{equation}
    -\Delta=\int_{0}^{r^*} u_{yy}(b) db.
\end{equation}
We expand this expression in the limit $a/r^* \ll 1$, neglecting terms quadratic in $a/r^*$ to find 
\begin{equation}
\Delta= -\frac{\zeta}{E \pi} \left(\ln \left(\frac{16 r^{*2}}{a^2}\right) + \frac{1}{2} (\pi -6 - \pi \nu) \right).
\end{equation}
This expression can be inverted using the Lambert W-function or product logarithm  (\cite{NIST:DLMF}, \textsection 4.13). Note that this function is multivalued for small indenter force and the $k=-1$ branch must be chosen for consistent results. 

\subsection{St. Venant-Kirchhoff model}
As described in Sec. IIIB of the main text, we simulate St. Venant-Kirchhoff materials surrounded by four circular indenters in FEniCS. We use the same mesh as in the hydrogel trials (approximately 30 vertices across the radius; see Fig. 3, left column). The strain-energy density function is
\begin{align}
    W&=G_0 u_{ij}^2+ \frac{\lambda}{2} u_{kk}^2=G_0 \left(u_{ij}^2+ \frac{\nu}{1-\nu} u_{kk}^2\right),\\
    u_{ij}&=\frac{1}{2} \left( \frac{\partial u_i}{\partial x_j} + \frac{\partial u_j}{\partial x_i} + \frac{\partial u_{k}}{\partial x_i} \frac{\partial u_k}{\partial x_j}\right),
\end{align}
where $G_0=\frac{Y}{2(1+\nu)}$ and $\lambda= \frac{Y \nu}{1-\nu^2}=\frac{2 G_0 \nu}{1-\nu}$ are Lam\'e coefficients, set to match the effective elastic properties of the hydrogel as described in ESI Sec. \ref{sec:elasticmoduli}, $u_i$ is a displacement in the $i$th direction, and indices run over $x$ and $y$. Displacements are defined relative to the zero stress reference configuration with radius $r^*$.

\begin{figure}[htp]
    \begin{center}
        \includegraphics[width=\textwidth]{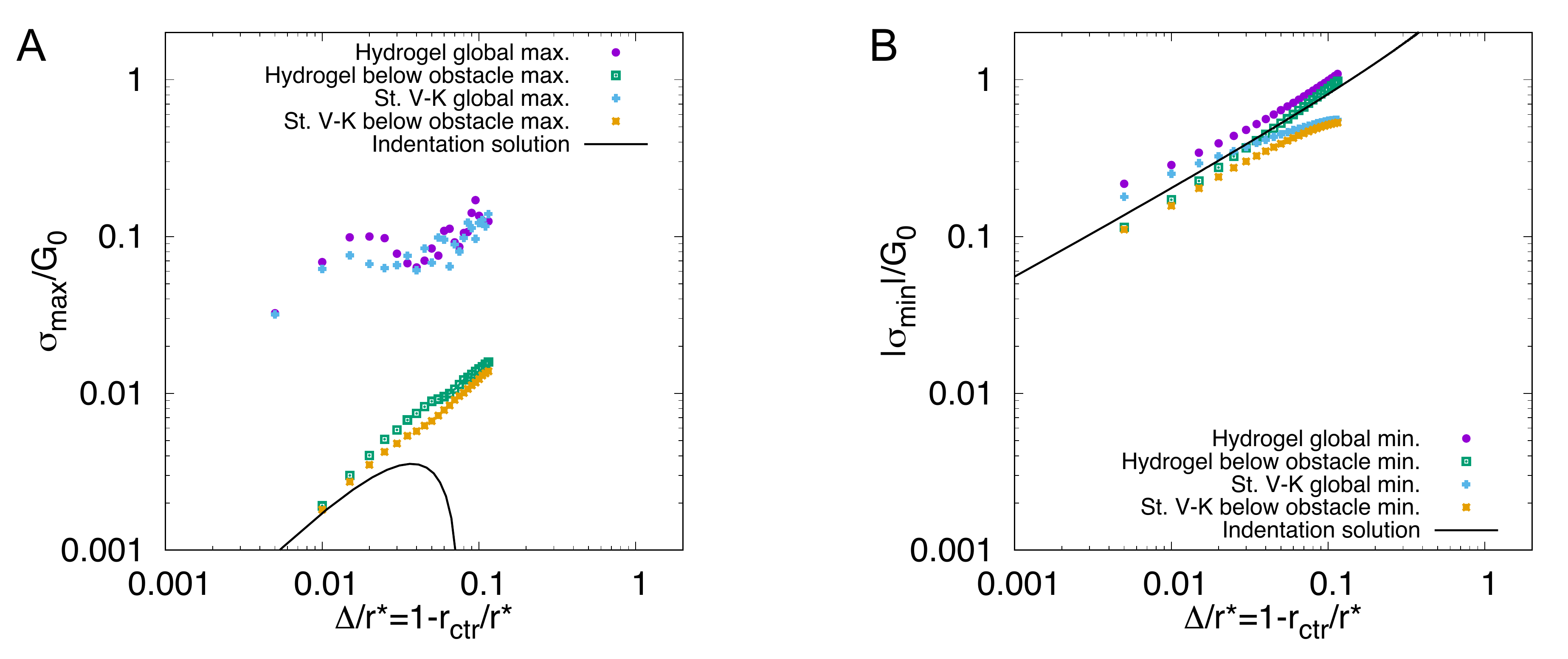}
\end{center}
    \caption{Comparison between maximum principal Cauchy tensile and compressive stresses in simulations with the nonlinear hydrogel model and St. Venant-Kirchhoff model. In (A), we observe that the St. Venant-Kirchhoff model produces approximately the same tensile stresses as the hydrogel model, implying that a geometric nonlinearity can explain tensile stresses up to $\sim \Delta/r^*=0.115$. In (B), we see that the St. Venant-Kirchhoff model displays compressive stresses that are lower than the hydrogel model, implying that a material nonlinearity is responsible for the additional compressive stress. Note that we plot both the maximum/minimum principal stress over the entire hydrogel domain (labeled global max./min.) as well as the maximum/minimum principal stress along the line connecting the hydrogel center to an obstacle center (labeled below obstacle max./min.).}
    \label{fig:tenscompr}
\end{figure}

\begin{figure}
    \begin{center}
        \includegraphics[width=\textwidth]{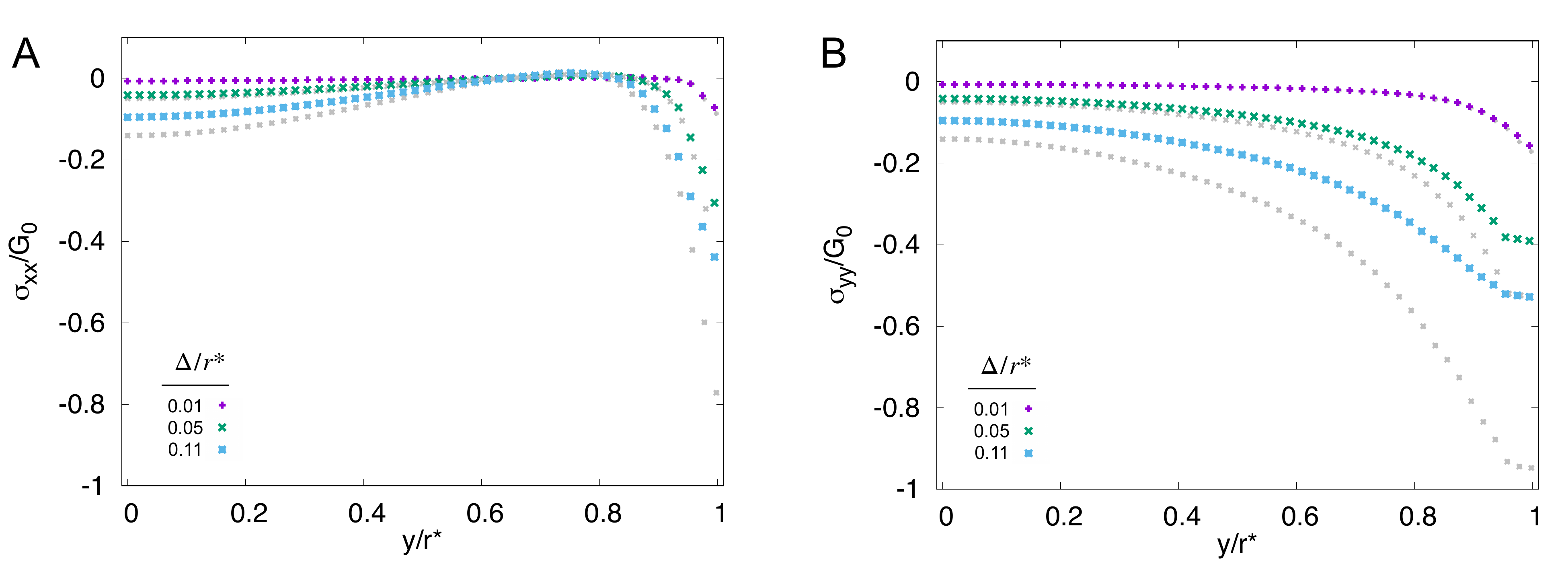}
\end{center}
    \caption{Comparison between Cauchy stress profiles in simulations with the nonlinear hydrogel model and the St. Venant-Kirchhoff elastic model, plotted against undeformed fully-swollen coordinates (corresponding to an unobstructed equilibrium state for the hydrogel). St. Venant-Kirchhoff results are shown with large colored markers, while hydrogel results are shown in grey. Data corresponding to the same value of $\Delta/r^*$ have the same marker shape.}
    \label{fig:stVprofiles}
\end{figure}

As the indenter displacement increases, St. Venant-Kirchhoff simulations can become unstable. For example, due to the linear constitutive law, there is a finite energetic cost for compressing material to a point \cite{sautter2022limitations}. To maintain stability when possible, we incrementally increase both indenter displacement and penalty strength. Nonetheless, we cannot simulate values of $\Delta/r^*>0.115$ for $r_{\rm{obs}}/r^*=0.3$. Results from these simulations prior to this threshold are shown in Fig. \ref{fig:tenscompr} and \ref{fig:stVprofiles}. 

As discussed in the main text, the St. Venant-Kirchhoff model is able to capture the maximum tensile stress reasonably well. However, the compressive stresses in the hydrogel model (and neo-Hookean model, see Fig. \ref{fig:nHprofiles}B) are significantly larger than those in the St. Venant-Kirchhoff model for large deformations.

\subsection{Maximum tensile stresses in weak confinement}\label{sec:maxmax}
\begin{figure}[htp]
    \begin{center}\includegraphics[width=0.7\textwidth]{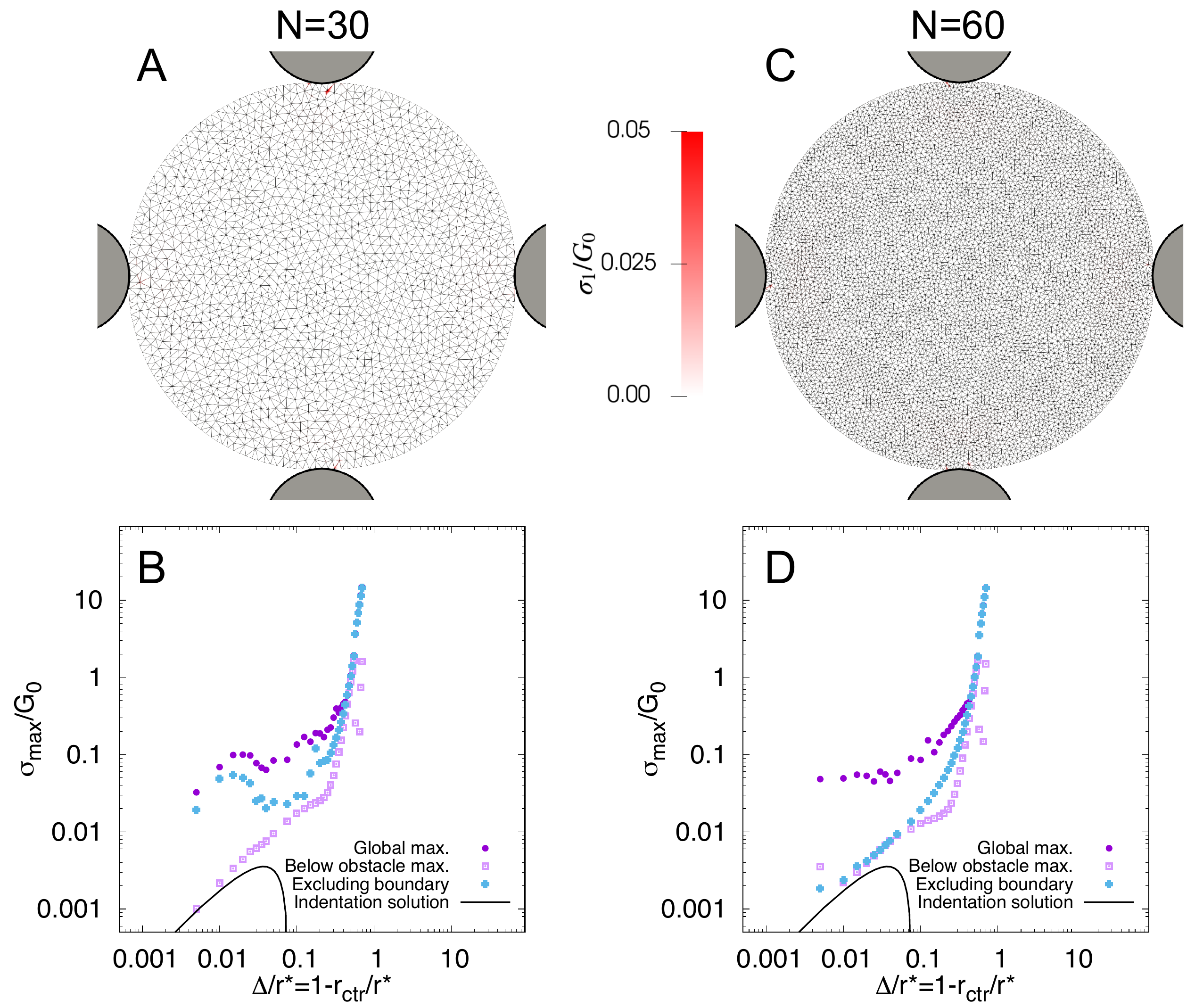}
\end{center}
    \caption{Large tensile stresses appear at isolated points near where the hydrogel loses contact with an obstacle. (A)~Positive values of the principal stresses for $\Delta/r^*=0.98, r_{\rm{obs}}/r^*=0.3$, plotted on top of the deformed hydrogel mesh and rescaled by the fully swollen shear modulus $G_0$. (B)~We compare the maximum principal stresses shown in Fig.~5 of the main text (purple circles) to the maximum principal stress excluding the boundary cells, taken by excluding cells at positions $\geq 95 \%$ of the hydrogel radius. Panels (C, D) show the same calculations at twice the resolution.}
    \label{fig:noise}
\end{figure}

In Figs. \ref{fig:nHmapping} and \ref{fig:tenscompr}, the global maximum of principal tensile stress is larger than the theoretical expectation at small indentation depths. These deviations are the result of isolated cells experiencing large tensile stresses on the boundary of the hydrogel where the material loses contact with the obstacle, and we consider it to be an unphysical effect originating with our finite mesh resolution. In Fig. \ref{fig:noise}, we show how the tension changes as we increase the resolution. At the relatively coarse resolution used for the simulations in this work with 30 vertices along the radius, the anomalous tension appears close to the boundary, while at higher resolutions it is more reliably located on the boundary itself. We compare the maximum principal stress excluding the boundary points (blue crosses in Fig. \ref{fig:noise}) to the maximum principal stress including the boundary and the maximum tensile stress beneath the obstacle. We find good agreement between the maximum stress excluding the boundary and the theoretical expectation at high resolution. However, since the maximum compression occurs close to the boundary as well, excluding these points systematically creates disagreement between theory and simulation for compressive stresses. Thus, in Fig. 3 we display the maximum and minimum stresses taken along a line connecting the obstacle center to the hydrogel center beneath the top obstacle, as decribed in \textit{Materials and methods}. 

\subsection{Symmetry-breaking instability}
\begin{figure}
    \centering
    \includegraphics[width=0.5\linewidth]{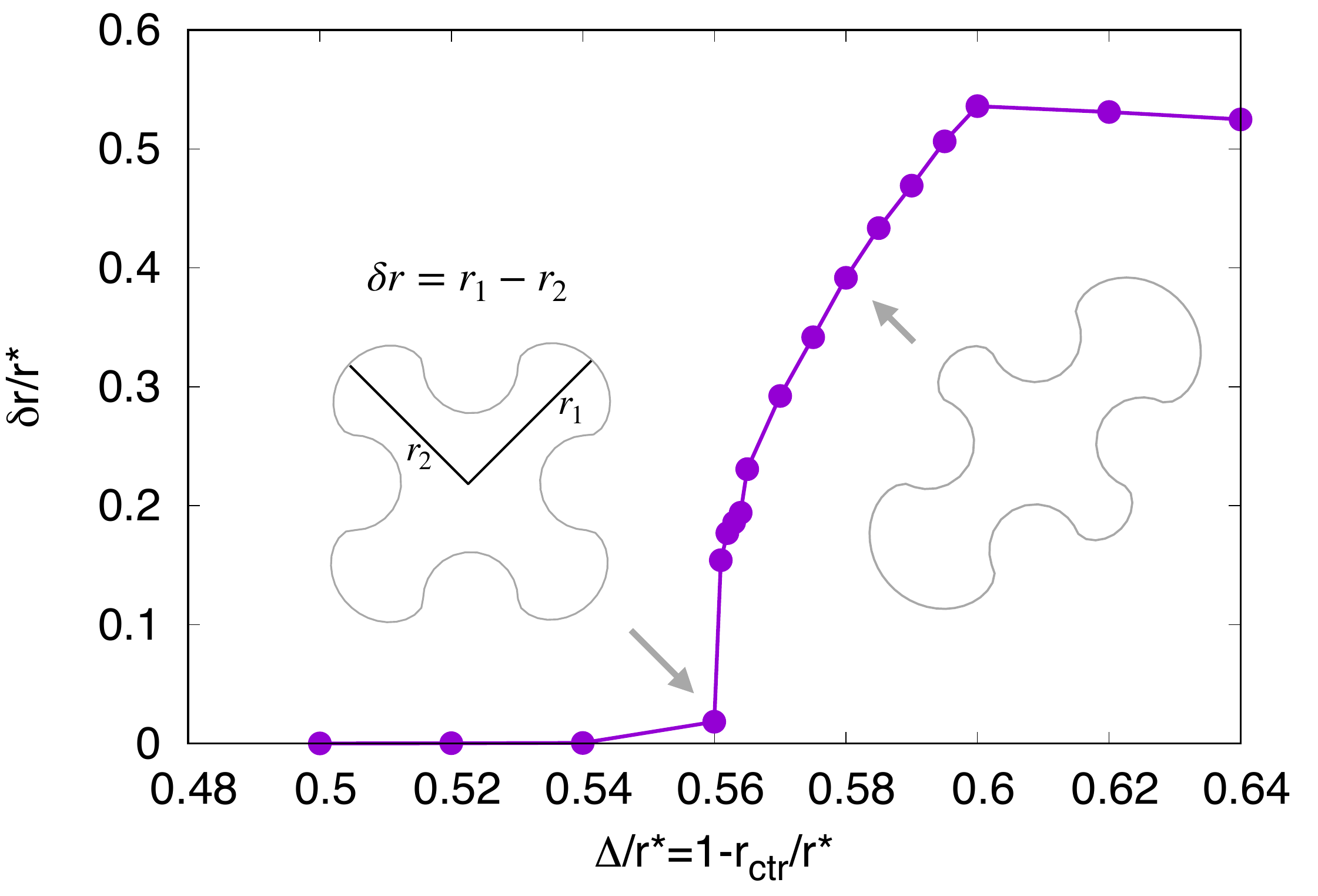}
    \caption{As $\Delta/r^*$ increases (the obstacles are brought closer together), the hydrogel begins to swell preferentially along a diagonal. The difference in the maximum distance from the origin of the top right and top left hydrogel lobe, $\delta r$, provides an order parameter for this transition. Insets show the boundary of the hydrogel at the indicated points (obstacles not pictured).}
    \label{fig:instability}
\end{figure}

When obstacles are very close together ($\Delta/r^* > 0.56$ for $r_{\rm{obs}}/r^* =0.3$), simulations show that the hydrogel disk swells primarily along a diagonal, rather than equally in all four pore spaces. This effect is quantified in Fig. \ref{fig:instability}.

As discussed in the main text, this instability can be understood as the hydrogel prioritizing an elliptical shape over a circular shape. In the weak confinement regime, deformations are relatively localized. Therefore, by symmetry, the hydrogel will swell evenly into all four pores. However, as the confinement increases and the deformation becomes global, maintaining symmetric swelling requires isotropic compression of the center of the hydrogel. Assuming uniform deformations and reasonable elastic parameters, it is less costly for the hydrogel to compress an amount $\Delta$ along a single axis, forming an ellipse, than to compress an amount $\Delta$ along two axes, forming a smaller circle.  For concreteness, we demonstrate this with a compressible neo-Hookean model (Eq. \ref{eq:compneohook}). Compressibility results from the migration of solvent molecules, as shown in ESI Sec. \ref{sec:elasticmoduli}. To transform a circle into a smaller circle, we impose stretches $\lambda_1=\lambda_2=1-\Delta/r^*$. The energy density is therefore
\begin{equation}
    W_{\rm{circle}}\propto 2\left( 1-\frac{\Delta}{r^*}\right)^2 - 2-2\ln\left[\left(1-\Delta/r^*\right)^2\right]+\frac{2 \nu}{(1-\nu)}\ln\left[\left(1-\Delta/r^*\right)^2\right]^2.
\end{equation}
To transform a circle into an ellipse, we impose $\lambda_1=1-\Delta/r^*, \lambda_2=1$. The energy for this deformation is
\begin{equation}
    W_{\rm{ellipse}}\propto \left( 1-\frac{\Delta}{r^*}\right)^2 - 1-2\ln\left[1-\Delta/r^*\right]+\frac{2 \nu}{(1-\nu)}\ln\left[1-\Delta/r^*\right]^2.
\end{equation}

We can expand both expressions in $\Delta/r^*$. At lowest order, both energies are quadratic in $\Delta/r^*$, with the circular deformation having a higher energy than the elliptical deformation as long as $\nu > -1/2$. We can also substitute in the effective Poisson's ratio used in simulations, $\nu=0.34$ (Eq. 2 of main text), to confirm that the elliptical deformation remains lower in energy as $\Delta/r^*$ increases to 1. 
%
 